\documentclass[a4paper,11pt]{article}
\pdfoutput=1 
\usepackage{jheppub} 
\usepackage[T1]{fontenc} 
\usepackage{braket}
\usepackage{amsmath}
\usepackage{amssymb}
\usepackage{mathrsfs}
\usepackage{xcolor}
\usepackage{graphicx}
\usepackage{comment}

\title{Gravitational form factors of a kink in $1+1$ dimensional $\phi^4$ model }

\author[a,b]{Hiroaki Ito}
\author[b,c,a]{Masakiyo Kitazawa}
\affiliation[a]{Department of Physics, Osaka University, Toyonaka, Osaka 560-0043, Japan
}
\affiliation[b]{Yukawa Institute for Theoretical Physics, Kyoto University, Kyoto 606-8502, Japan}
\affiliation[c]{J-PARC Branch, KEK Theory Center, Institute of Particle and Nuclear Studies, Tokai, Ibaraki, KEK, 319-1106, Japan}

\emailAdd{ito@kern.phys.sci.osaka-u.ac.jp}
\emailAdd{kitazawa@yukawa.kyoto-u.ac.jp}
\preprint{YITP-23-18, J-PARC-TH-0284}

\abstract{We calculate the one-loop correction to the distribution of energy-momentum tensor around a kink in $1+1$ dimensional $\phi^4$ model. We employ the collective coordinate method to eliminate the zero mode that gives rise to infrared divergence. The ultraviolet divergences are removed by vacuum subtraction and mass renormalization. We obtain an analytic result that is finite and satisfies the momentum conservation. The total energy of the kink obtained from the spatial integral of energy density reproduces the known result. Our result obtained on a finite space has a spatially-uniform term that is inversely proportional to the spatial length.}

\begin{document} 
\maketitle
\flushbottom

\section{Introduction}
\label{sec:intro}

Energy-momentum tensor (EMT), $T^{\mu\nu}(x)$, is a fundamental observable in physics that is closely related to space-time symmetry. EMT plays indispensable roles in classical and quantum field theories for various purposes; for example, its individual components, energy and momentum densities, and stress tensor, are basic quantities having definite physical meanings.

Recently, there has been remarkable progress in using the EMT operator for investigating localized systems in quantum field theory. An example is the experimental investigations of the gravitational form factors (GFF) of hadrons~\cite{Ji:1996nm,Sawada:2016mao,Kumano:2017lhr,Burkert:2018bqq}, that is the matrix element of the EMT operator~\cite{Ji:1996ek,Hudson:2017xug,Polyakov:2018zvc,Hatta:2018sqd,Freese:2019bhb,Fujita:2022jus,Ding:2022ows,Tanaka:2022wzy}. The GFF in the coordinate space represent the mechanical structure of hadrons~\cite{Polyakov:2002yz,Polyakov:2018zvc}, and provide us with novel insights into the hadron structure. Their detailed study is one of the central goals of the Electron-Ion Collider (EIC)~\cite{Burkert:2022hjz}, and precise experimental data will be provided in the future. The measurement of the GFF on the lattice is also ongoing~\cite{Hagler:2009ni,Shanahan:2018nnv}.

Another progress has been made in the numerical analysis of static-quark systems in lattice gauge theory. Thanks to an efficient method to measure the expectation value of the EMT operator on the lattice~\cite{Suzuki:2013gza,Asakawa:2013laa,Makino:2014taa,Kitazawa:2016dsl,Iritani:2018idk,Taniguchi:2016ofw,Taniguchi:2020mgg} based on the gradient flow~\cite{Luscher:2011bx,Narayanan:2006rf}, detailed analysis of the local distribution of EMT in various non-uniform and non-isotropic systems has been realized~\cite{Yanagihara:2018qqg,Kitazawa:2019otp,Yanagihara:2020tvs}. In particular, the numerical result of the static quark--anti-quark ($Q\bar Q$) system~\cite{Yanagihara:2018qqg} has revealed the formation of the flux tube and its mechanical structure in terms of the gauge-invariant observable.

In these localized systems, quantum effects should play crucial roles in determining the EMT distribution. For example, in the $Q\bar Q$ system it is known that the width of the flux tube becomes larger with increasing the $Q\bar Q$ distance due to quantum string vibrations~\cite{Luscher:1980ac,Luscher:1980iy,Gliozzi:2010zv,Cardoso:2013lla}. Its importance is also suggested from the comparison of the lattice result in Ref.~\cite{Yanagihara:2018qqg} with the classical EMT distribution around the flux tube in the dual superconductor model~\cite{Yanagihara:2019foh}. The pressure anisotropy induced by boundaries also arises from purely quantum effects~\cite{Brown:1969na,Kitazawa:2019otp}. To understand these experimental and numerical results, therefore, investigations of the quantum effects on the EMT distribution are inevitable.

In the present study, as a trial of such investigations, we focus on the kink in the $1+1$ dimensional scalar $\phi^4$ theory and calculate the EMT distribution around it incorporating quantum effects to one-loop order. The kink, which is also called the soliton, is a localized and stable classical solution in this theory that connects two degenerate vacua~\cite{Rajaraman:1982is}. Its properties and applications have been discussed actively more than half century~\cite{Dashen:1974cj,Dashen:1975hd,Coleman:1974bu,Jackiw:1977yn,Rajaraman:1982is,Miyashita:1983rkl,Yamagishi:1984zv,Shifman:1998zy,Goldhaber:2001rp,Alonso-Izquierdo:2011hmo,Papageorgakis:2014dma,Melnikov:2020ret,Evslin:2021vgk,Martin:2022pri,Mukhopadhyay:2021wmu,Wheater:2022led,Evslin:2022opz}. However, its EMT distribution at the quantum level has not been understood well to the best of the authors' knowledge. As for related studies, the quantum correction to the {\it total} energy of the kink has been calculated at one-loop order in the renowned paper by Dashen, et al.~\cite{Dashen:1974cj}, and the result has been confirmed in many literature~\cite{Rajaraman:1982is,Boya:1989db,Rebhan:2002uk,Bordag:2002dg,Goldhaber:2004kn,Gousheh:2012qu,Graham:2022rqk}. Also, there are several attempts to calculate the energy {\it density}~\cite{Goldhaber:2001rp,Martin:2022pri}, i.e. the expectation value of $T^{00}(x)$\footnote{Also, the mean-square radius of the energy density has been evaluated in Ref.~\cite{Wheater:2022led}.}. However, these studies have not investigated the spatial component $T^{11}(x)$. In the present study, we calculate all components simultaneously. We show that our result satisfies the momentum conservation. However, the expectation value of $T^{00}(x)$ does not agree with any of those in Refs.~\cite{Goldhaber:2001rp,Martin:2022pri}, while the spatial integral of $T^{00}(x)$ reproduces the total energy in Ref.~\cite{Dashen:1974cj} in all the results.

In this analysis, we face a difficulty arising from the zero mode in the fluctuations around the classical solution, which physically represents the space translation of the kink.
The zero mode causes an infrared divergence in the perturbative expansion. It also brings about a conceptual difficulty in the definition of the EMT distribution around the kink in quantum systems, since the location of the kink is not fixed in the quantum ground state. It is known that these problems are resolved by employing the collective coordinate method (CCM)~\cite{Gervais:1974dc,Gervais:1975pa,Tomboulis:1975gf,Christ:1975wt}, in which the zero mode is eliminated by promoting the coordinate of the kink to a dynamical variable. The CCM also allows us to define the EMT distribution of the kink around its center-of-mass frame, which is the Fourier transform of the GFF~\cite{Rajaraman:1982is,Polyakov:2018zvc}. We will discuss these issues in Sec.~\ref{sec:CCM}.

The analysis at one-loop order also has ultraviolet (UV) divergences. We eliminate them in two steps; vacuum subtraction and mass renormalization. For the former, we employ the same procedure as in Ref.~\cite{Dashen:1974cj}, which is named the mode-number cutoff (MNC) scheme~\cite{Rebhan:1997iv}. In this method, the subtraction between the kink and vacuum sectors is performed in a finite system of length $L$ assuming that each sector has the same mode numbers. The result after the vacuum subtraction is still logarithmically divergent, which can be removed by mass renormalization.

We show that our result of $T^{00}(x)$ and $T^{11}(x)$ obtained at the spatial length $L$ has a constant term proportional to $1/L$. This term has a finite contribution to the total energy in the $L\to\infty$ limit, while it vanishes in the local EMT distribution. The total energy in Ref.~\cite{Dashen:1974cj} is reproduced including this contribution. This result means that the integral of the local EMT distribution defined in the $L\to\infty$ limit is not consistent with the result in Ref.~\cite{Dashen:1974cj}. 

This paper is organized as follows. In the next section we introduce the $\phi^4$ theory and its kink solution, and summarize their basic properties. In Sec.~\ref{sec:CCM} we give a brief review of the CCM. The expectation values of EMT around the kink are then calculated in Sec.~\ref{sec:perturbative}, and the final result and its properties are discussed in Sec. \ref{sec:result}. The final section is devoted to a summary and outlook. The topological charge density is calculated in App.~\ref{sec:topological}. In App.~\ref{app:Regularization}, App.~\ref{sec:MNC} and App.~\ref{app:tadpole}, specific topics on the mass renormalization, vacuum subtraction based on the MNC, and analysis of the tadpole diagram, respectively, will be discussed. In App.~\ref{sec:LMR}, we discuss the analyses in Refs.~\cite{Goldhaber:2001rp,Martin:2022pri}.

\section{Model}
\label{sec:model}

We employ the real-scalar $\phi^4$ theory in a $1+1$ dimensional system, whose Lagrangian density is given by 
\begin{align}
    {\cal L} = \frac{1}{2}\partial_\mu \phi \partial^\mu \phi - U(\phi) , 
    \label{eq:L}
\end{align}
with the potential term
\begin{align}
U(\phi) = \frac{\lambda}{4}\left( \phi^2 -v^2 \right )^2 
= -\frac12 m^2 \phi^2 + \frac\lambda4 \phi^4 + \frac{\lambda v^4}4 ,
\label{eq:U}
\end{align}
where $\phi=\phi(x)$ is the real scalar field. The potential $U(\phi)$ has two degenerate minima at $\phi=\pm v$ with $v^2=m^2/\lambda$.

\subsection{Classical solutions}

The classical equation of motion (EoM) of this theory is given by
\begin{align}
    \partial_0^2 \phi - \partial_1^2 \phi + \frac{dU}{d\phi}
    = \partial_0^2 \phi - \partial_1^2 \phi - m^2 \phi + \lambda \phi^3 
    = 0 .
    \label{eq:EoM}
\end{align}
Since $U(\phi)$ has minima at $\phi=\pm v$, 
\begin{align}
    \phi_{\rm vac}(x) = \pm v = \pm \frac{m}{\lambda^{1/2}} ,
    \label{eq:vac}
\end{align}
are static solutions of Eq.~(\ref{eq:EoM}). We refer to these trivial solutions as the vacuum.

The EoM~(\ref{eq:EoM}) has other static solutions called the kink and anti-kink, 
\begin{align}
    \phi_{\rm kink}(x;X)= \pm \frac{m}{\lambda^{1/2}} \tanh \frac{m(x-X)}{\sqrt{2}},
    \label{eq:kink}
\end{align} 
where $X$ is a free parameter that represents the position of the kink. As Eq.~(\ref{eq:kink}) behaves $\phi_{\rm kink}(x;X)\to\pm v$ in the limit $x\to\infty$ or $x\to-\infty$, the (anti-)kink solution connects two vacua in Eq.~(\ref{eq:vac}). 

The EMT in this theory is given by the Noether current as
\begin{align}
    T^{\mu\nu}(x) = (\partial^\mu \phi)(\partial^\nu \phi )
    - \frac12 g^{\mu\nu} (\partial^\rho\phi)(\partial_\rho\phi) + g^{\mu\nu} U(\phi) .
    \label{eq:Tmunu}
\end{align}
Substituting Eqs.~(\ref{eq:vac}) and (\ref{eq:kink}) into Eq.~(\ref{eq:Tmunu}), one finds that $T^{\mu\nu}(x)=0$ for the vacuum and
\begin{align}
    T^{00}_{\rm kink}(x)
    =\frac{m^4}{2\lambda}{\rm sech}^4 \frac{m(x-X)}{\sqrt{2}} ,
    \qquad
    T^{01}_{\rm kink}(x) = T^{11}_{\rm kink}(x) = 0,
    \label{eq:Tmunu_kink}
\end{align}
for the kink with ${\rm sech}x=1/\cosh x$. By integrating $T^{00}_{\rm kink}(x)$, we obtain the total energy 
\begin{align}
    E_{\rm kink}= \int dx T^{00}_{\rm kink}(x) = \frac{2\sqrt{2}m^3}{3\lambda} .
    \label{eq:Ekink}
\end{align}

In the following, we evaluate the quantum correction to Eq.~(\ref{eq:Tmunu_kink}) to the leading order of perturbative expansion with respect to $\lambda$; the dimensionless expansion parameter is $\lambda/m^2$, or $\lambda\hbar/m^2$ if $\hbar$ is explicitly shown. Since Eq.~(\ref{eq:Tmunu_kink}) is of order $\lambda^{-1}$, the leading-order correction to it is at order $\lambda^0$. We also note that $\phi_{\rm kink}(x)$ is of order $\lambda^{-1/2}$ as in Eq.~(\ref{eq:kink}).

One can also define the topological current~\cite{Rajaraman:1982is} 
\begin{align}
\label{eq:topologicalcurrent}
  j^\mu(x)=\frac{\lambda^{1/2}}{2m}\epsilon^{\mu \nu }\partial_{\nu }\phi(x),
\end{align}
that satisfies the current conservation $\partial_{\mu}j^{\mu}=0$, where $\epsilon^{\mu \nu }$ is the anti-symmetric tensor. From Eq.~(\ref{eq:kink}) one has
\begin{align}
 j^0_{\rm kink}(x) =\frac{\lambda^{1/2}}{2m}\partial_{1} \phi_{\rm kink }(x;X)
 =\pm \frac{m}{2}{\rm sech}^2 \frac{m(x-X)}{\sqrt{2}}   .
 \label{eq:topol}
\end{align}
The topological charge $Q$ is given by the spatial integral of $j^0$;
\begin{align}
 Q= \int^{\infty}_{-\infty} dx j^0 (x)
 = \frac{\lambda^{1/2}}{2m}  [\phi(\infty;X)-\phi(-\infty;X)] =
    \begin{cases}
       \pm1 & (\mbox{kink/anti-kink}),\\
        0 &  (\mbox{vacuum}).
    \end{cases}
\end{align}
In App.~\ref{sec:topological}, we calculate the quantum correction to Eq.~(\ref{eq:topol}).

\subsection{Expansion around the classical solutions}

To calculate the quantum correction to Eq.~(\ref{eq:Tmunu_kink}), we expand the field $\phi(x,t)$ around the classical solutions as 
\begin{align}
    \phi(x,t) &=  v + \chi(x,t) , 
    \label{eq:chi}
    \\
    \phi(x,t) &= \phi_{\rm kink}(x;X) + \eta(x,t) , 
    \label{eq:eta}
\end{align}
where we take the positive sign in Eq.~(\ref{eq:kink}) in the following. The action is written in terms of $\chi(x,t)$ and $\eta(x,t)$ as
\begin{align}
    S&= \int dx^2 {\cal L} 
    \nonumber \\
    &= S_{\rm vac} + \int dx^2 \Big[\frac{1}{2}(\partial_0 \chi)^2 -\frac{1}{2}(\partial_1 \chi)^2- m^2\chi^2 - \lambda^{1/2}m\chi^3-\frac{\lambda}{4}\chi^4 \Big] 
    \label{eq:S(chi)} \\
    &= S_{\rm kink}+\int dx^2 \Big[ \frac{1}{2}(\partial_0 \eta)^2 -\frac{1}{2}(\partial_1 \eta)^2- \frac{\lambda}{2}
    \big(3\phi^2_{\rm kink}-v^2 \big)\eta^2 -\lambda\phi_{\rm kink}\eta^3-\frac{\lambda}{4}\eta^4 \Big] ,
    \label{eq:S(eta)}
\end{align}
where $S_{\rm vac} = S[v]$ and $S_{\rm kink} = S[\phi_{\rm kink}(x;X)]$ are the classical action of each sector. We note that terms linear in $\chi(x,t)$ or $\eta(x,t)$ are eliminated by the partial integral and the EoM~(\ref{eq:EoM}).

The quadratic terms in Eq.~(\ref{eq:S(eta)}),
\begin{align}
    -\frac12 \int d^2 x \eta (  \partial_0^2 + \Delta ) \eta ,
    \qquad
    \Delta = -\partial_1^2 + \lambda \big( 3 \phi_{\rm kink}^2 - v^2 \big) ,
\end{align}
are diagonalized by solving the eigenequation
\begin{align}
\Delta \psi_l(x) = \omega^2_l \psi_l(x) .
\label{eq:eigen_eq}
\end{align}
The analytic solution of Eq.~(\ref{eq:eigen_eq}) is known as~\cite{Morse-Feshbach:1953}
\begin{align}
\omega^2_0&=0,&\psi_0(x)=& {\rm sech}^2 \frac{mx}{\sqrt{2}} \sim \partial_1 \phi_{\rm kink}(x;0) ,
\label{eq:psi0}\\
\omega^2_1&=\frac{3}{2}m^2,&\psi_1(x)=& \sinh{\frac{mx}{\sqrt{2}}} {\rm sech}^2 \frac{mx}{\sqrt{2}} ,
\label{eq:psi1}\\
\omega^2_q&=q^2+2m^2,&\psi_q(x)=&e^{iqx}\Big(3\tanh^2\frac{mx}{\sqrt{2}}-1-\frac{2}{m^2}q^2-3\sqrt{2}i\frac{q}{m}\tanh\frac{mx}{\sqrt{2}}\Big),
\label{eq:psiq}
\end{align}
for $X=0$. Here, $\psi_0(x)$ and $\psi_1(x)$ are discrete modes, while $\psi_q(x)$ for real number $q$ form a continuous spectrum. $\psi_0(x)$ is proportional to $\partial_1\phi_{\rm kink}(x;0)$ and represents the space translation of the kink. It thus is called the translational mode. This mode is interpreted as the Nambu-Goldstone mode associated with the violation of translational invariance due to the existence of the kink. 
The continuous modes $\psi_q(x)$ have an asymptotic behaviour 
\begin{align}
    \psi_q(x)\xrightarrow[x \to \pm \infty]{} C \exp \Big(iqx\pm  \frac{i}{2}\delta_{p}(q)\Big) ,
    \label{eq:psiq->}
\end{align}
with a constant $C$ and the phase shift
\begin{align}
    \delta_{p}(q)=-2\arctan\frac{3\sqrt{2}mq}{2m^2-2q^2} .
    \label{eq:delta_p}
\end{align}
The argument of $\arctan$ diverges at $q=\pm m$, which means that the phase shift crosses $\delta_p(q)=\pm\pi$ there. Requiring $\delta_p(0)=0$, to make $\delta_p(q)$ continuous we obtain\footnote{ If we take phase shift to behave $\delta_{p}(q) \to 0$ for $|q|\to \infty $, $\delta_{p}(q)$ becomes discontinuous at $q=0$. This choice of the phase shift leads to the same result as discussed in Ref.~\cite{Nastase:1998sy}.} ~\cite{Rebhan:1997iv}
\begin{align}
    \delta_p(q) \xrightarrow[q\to\pm\infty]{} \mp 2\pi \pm 3\sqrt{2}\frac mq .
    \label{eq:delta_p->}
\end{align}

For later use, we introduce the normalized eigenmodes $\bar\psi_l(x)$ where $l$ represents all the eigenmodes. Because the following analysis is mainly performed in a finite system of length $L$ where the continuous modes are discretized, we impose the orthogonality condition
\begin{align}
    \int_{-L/2}^{L/2}dx \bar\psi_{l_1}^*(x) \bar\psi_{l_2}(x) = \delta_{l_1l_2}.
    \label{eq:ortho}
\end{align}
For the discrete modes $l=0,1$, we obtain 
\begin{align}
    \bar\psi_0(x) = \Big ( \frac{3m}{4\sqrt{2}}  \Big )^{1/2}\psi_0(x) ,
    \qquad
    \bar\psi_1(x) = \Big ( \frac{3m}{2\sqrt{2}}  \Big )^{1/2} \psi_1(x) ,
    \label{eq:tildepsi}
\end{align}
where the effect of finite $L$ is exponentially suppressed for $mL\gg1$.
For the continuous modes, using 
\begin{align}
    \label{eq:psiq2}
    |\psi_q|^2 
    =& \Big( 3\tanh^2\frac{mx}{\sqrt2} - 1 -  \frac{2q^2}{m^2}  \Big)^2
    + 18  \frac{q^2}{m^2}   \tanh^2\frac{mx}{\sqrt2}
    \nonumber \\
    =& \frac{2}{m^4}(2q^2+m^2)(q^2+2m^2) - \frac{3}{m^2}(2q^2+m^2)\psi_0^2 - \frac{6}{m^2}(q^2+2m^2)\psi_1^2 ,
\end{align}
the normalization constant is calculated to be
\begin{align}
    N_q =& \int^{L/2}_{-L/2} dx |\psi_q|^2 = \frac{2L}{m^4}(2q^2+m^2)(q^2+2m^2) -\frac{12\sqrt{2}}{m^3} (q^2+m^2)
    \notag \\
    =& \frac{2L}{m^4}(2q^2+m^2)(q^2+2m^2) \Big( 1 + \frac1L \delta_p'(q) \Big),
\end{align}
which gives $\bar\psi_q(x) = \psi_q(x)/\sqrt{N_q}$ with 
\begin{align}
    \delta_p'(q) = \frac{d\delta_p(q)}{dq} 
    = -\frac{6\sqrt{2} m(q^2+m^2)}{(2q^2+m^2)(q^2+2m^2)} .
    \label{eq:delta_p'}
\end{align}

For the boundary conditions (BC), we impose the anti-periodic BC (APBC)
\begin{align}
    \phi(x+L)=-\phi(x),
    \label{eq:APBC}
\end{align}
unless otherwise stated, since this choice of the BC conforms to Eq.~(\ref{eq:kink}). The effect of the boundary in the analysis of the total energy has been discussed in the literature~\cite{Rebhan:1997iv,Goldhaber:2000ab}. Their conclusion is that the total energy does not depend on the choice of the BC. Later, we will argue that the APBC removes a divergence that appears in the calculation of a tadpole diagram most naturally. From Eq.~(\ref{eq:APBC}) that means $\eta(x+L)=-\eta(x)$ and Eq.~(\ref{eq:psiq->}), the values of $q$ are restricted to discrete ones satisfying 
\begin{align}
    L q_n + \delta_{p}(q_n) = (2n+1)\pi,
    \label{eq:q_n}
\end{align}
for $L\to\infty$ with integer $n$.

Using the normalized eigenfunctions, $\eta(x)$ is represented as
\begin{align}
    \eta(x) = c_0 \bar\psi_0(x) + c_1 \bar\psi_1(x) +\sum_n c_{q_n} \bar\psi_{q_n}(x) 
    = \sum_l c_l \bar\psi_l(x),
    \label{eq:eta=cpsi}
\end{align}
where the sum on the far right-hand side runs over $l=0$, $1$ and $q_n$.
The quadratic Hamiltonian is expressed in terms of $c_l$ as
\begin{align}
    H = \frac12 \sum_l \omega_l c_l^2 .
\end{align}

For the vacuum sector, the eigenmodes are discretized as
\begin{align}
    \varphi_n(x) = e^{ik_nx},
    \qquad
    k_n = \frac{(2n+1)\pi }L,
    \label{eq:k_n}
\end{align}
with the APBC $\chi(x+L)= -\chi(x)$~\footnote{We impose the same APBC for the vacuum sector. This choice makes the vacuum subtraction transparent as discussed in Appendix~\ref{sec:MNC}. However, the APBC would be inconsistent with the periodicity of the classical solution in the vacuum sector. For further discussion, see Appendix~\ref{sec:MNC}.}.

Substituting Eqs.~(\ref{eq:chi}) and (\ref{eq:eta}) into Eq.~(\ref{eq:Tmunu}), EMT is rewritten as 
\begin{align}
T^{00} 
&= \frac{1}{2}(\partial_0 \chi)^2 +\frac{1}{2}(\partial_1 \chi)^2 +m^2\chi^2+ \mathcal{O}(\lambda^{1/2})  ,
\label{eq:T00chi}
\\
T^{11}
&=  \frac{1}{2}(\partial_0 \chi)^2 +\frac{1}{2}(\partial_1 \chi)^2 -m^2\chi^2+ \mathcal{O}(\lambda^{1/2})  ,
\label{eq:T11chi}
\\
T^{01} &= -(\partial_0 \chi)(\partial_1 \chi) ,
\end{align}
for the vacuum sector and
\begin{align}
    T^{00} =& T^{00}_{\rm kink} +\frac{1}{2}(\partial_0 \eta)^2 +\frac{1}{2}(\partial_1 \eta)^2 +(\partial_1 \phi_{\rm kink})(\partial_1 \eta)
    \notag \\
    &+\lambda\phi_{\rm kink}(\phi^2_{\rm kink}-v^2)\eta  
    +\frac{\lambda}{2}(3\phi^2_{\rm kink}-v^2)\eta^2+O(\lambda^{1/2}) ,
    \label{eq:T00eta}
\\
T^{11} =& +\frac{1}{2}(\partial_0 \eta)^2 +\frac{1}{2}(\partial_1 \eta)^2 -(\partial_1 \phi_{\rm kink})(\partial_1 \eta)-\lambda\phi_{\rm kink}(\phi^2_{\rm kink}-v^2)\eta  \notag \\
&-\frac{\lambda}{2}(3\phi^2_{\rm kink}-v^2)\eta^2+O(\lambda^{1/2}) ,
\label{eq:T11eta}
\\
T^{01} =&-(\partial_0 \eta)(\partial_1 \eta)-(\partial_0 \eta)(\partial_1 \phi_{\rm kink}) ,
\end{align}
for the kink sector, where we omitted higher order terms that are negligible to order $\lambda^0$. We note that Eqs.~(\ref{eq:T00eta}) and~(\ref{eq:T11eta}) have linear terms in $\eta(x)$, while such terms do not appear in the action~(\ref{eq:S(eta)}) as they are eliminated by the partial integral and the EoM. We will see later that these linear terms calculated from the tadpole diagrams have nonzero contributions.

\section{Collective-coordinate method}
\label{sec:CCM}

In the perturbative analysis, the zero mode in Eq.~(\ref{eq:psi0}) leads to an infrared divergence. The appearance of the zero mode is related to the fact that the kink position $X$ is arbitrary and the translation of the kink requires zero energy. The zero mode also causes another conceptual difficulty. In the ground state of this system in quantum theory, the value of $X$ is not fixed, but the ground state is the eigenstate of the conjugate momentum of $X$. Hence, the expectation value of EMT is uniform in space in the ground state. To obtain a non-trivial result, one has to introduce a quantum expectation value with fixed $X$.

It is known that these problems are resolved by employing a procedure called the collective-coordinate method (CCM)~\cite{Gervais:1974dc,Gervais:1975pa,Tomboulis:1975gf}. In the CCM, the perturbative analysis is performed by eliminating the zero mode in place of the promotion of $X$ to a dynamical variable. In this section, we give a brief review of the CCM to make the manuscript self-contained. The CCM has been formulated by various methods, such as the canonical and the path-integral formalisms, which give the same result~\cite{Gervais:1974dc,Gervais:1975pa,Tomboulis:1975gf,Christ:1975wt,Rajaraman:1982is}. In this section, we illustrate the CCM based on Refs.~\cite{Tomboulis:1975gf,Christ:1975wt}. See also Sec.~8 of Ref.~\cite{Rajaraman:1982is}.

\subsection{Canonical transformation}

Let us start from the classical system described by the Lagrangian~(\ref{eq:L}). There are various choices for a set of dynamical variables to describe the system; in addition to the original field $\phi(x,t)$, one can choose $\eta(x,t)$ in Eq.~(\ref{eq:eta}), or $c_l$ in Eq.~(\ref{eq:eta=cpsi}).

Now, let us rewrite $\phi(x,t)$ as 
\begin{align}
    \phi(x,t) = \phi_{\rm kink}(x;X(t)) + \tilde\eta(x-X(t),t) ,
    \label{eq:phi(x-X)}
\end{align}
and regard $X(t)$ as a dynamical variable. Since this causes redundancy in the degrees of freedom, we impose a constraint on $\tilde\eta(x,t)$
\begin{align}
    \int dx \tilde\eta(x,t) \bar\psi_0(x) = 0 .
    \label{eq:int_eta}
\end{align}
This constraint means that the variable $c_0$ in Eq.~(\ref{eq:eta=cpsi}), i.e. the zero mode, is removed and $\tilde\eta(x,t)$ is given by
\begin{align}
    \tilde\eta(x,t) = \sum_{l\ne0} c_l(t) \bar\psi_l(x) .
    \label{eq:tildeeta}
\end{align}
The basic idea of the CCM is to describe the system using the set of variables $X(t)$ and $\tilde\eta(x,t)$, or equivalently $X(t)$ and $c_l(t)$ for $l\ne0$.

The Hamiltonian of the system is represented in terms of the new variables by canonical transformation. For this we introduce the conjugate momenta of $X(t)$ and $\tilde\eta(x,t)$,
\begin{align}
    P(t) = \frac{\partial L}{\partial (\partial_0X)}, 
    \quad
    \tilde\pi(x,t) = \frac{\delta L}{\delta (\partial_0\tilde\eta)} ,
\end{align}
where $L=\int dx{\cal L}$ is the Lagrangian. The conjugate field $\tilde\pi(x,t)$ can also be defined as
\begin{align}
    \tilde\pi(x,t) = \sum_{l\ne0} \gamma_l(t) \bar\psi_l(x) ,
    \label{eq:tildepi}
\end{align}
with $\gamma_l=(\partial L)/(\partial(\partial_0c_l))$ being the canonical conjugate of $c_l$. In any case, $\tilde\pi(x,t)$ also satisfies the orthogonality condition
\begin{align}    
    \int dx \tilde\pi(x) \bar\psi_0(x) = 0 .
    \label{eq:int_pi}
\end{align}
It is found that the conjugate of the original field $\pi=\partial{\cal L}/\partial (\partial_0\phi)$ is given by
\begin{align}
    \pi(x,t) = \tilde\pi(x-X,t) - \frac{P(t)+\int dx \tilde\pi \partial_1\tilde\eta}{E_{\rm kink}^{1/2}(1+\xi/E_{\rm kink}^{1/2})} \bar\psi_0(x-X),
    \label{eq:pi(x-X)}
\end{align}
with $\xi=\int dx (\partial_1\tilde\eta(x))\bar\psi_0(x)$.

The variables $X(t)$ and $P(t)$ satisfies $\{X,P\}=1$, where $\{\cdot,\cdot\}$ is the Poisson bracket in this subsection. The Poisson bracket of $\tilde\eta(x,t)$ and $\tilde\pi(y,t)$ is given by
\begin{align}
    \{\tilde\eta(x), \tilde\pi(y)\} 
    =\sum_{l \neq 0} \bar\psi_l(x) \bar\psi^{*}_l(y)
    = \delta(x-y) - \bar\psi_0(x)\bar\psi_0(y) ,
    \label{eq:[eta,pi]}
\end{align}
due to the constraints~(\ref{eq:int_eta}) and (\ref{eq:int_pi}). The deviation from the delta function in Eq.~(\ref{eq:[eta,pi]}) is understood as the Poisson bracket in constrained systems~\cite{Dirac:1964}. These Poisson brackets and Eq.~(\ref{eq:pi(x-X)}) give 
\begin{align}
    \{\phi(x),\pi(y)\} = \delta(x-y).
    \label{eq:[phi,pi]}
\end{align}

In terms of $X$, $P$, $\tilde\eta$, and $\tilde\pi$, the Hamiltonian of the system is written as 
\begin{align}
    H = E_{\rm kink} + \frac1{2E_{\rm kink}}\frac{(P+\int dx \tilde\pi \partial_1\tilde\eta)^2}{\big(1+\xi / E_{\rm kink}^{1/2}\big)^2} + \tilde{H},
    \label{eq:H}
\end{align}
with
\begin{align}
    \tilde{H} =& \int dx \tilde{\cal H}(x-X),
    \label{eq:tildeH}
    \\
    \tilde{\cal H}(x) =& \frac12 \tilde\pi^2(x) + \frac12 (\partial_1\tilde\eta(x))^2 + U(\phi_{\rm kink}(x;0)+\tilde\eta(x) )-U(\phi_{\rm kink}(x;0) ) .
    \label{eq:tildeH(x)}
\end{align}
In Eq.~(\ref{eq:H}), the first term $E_{\rm kink}$ represents the classical energy of the kink~(\ref{eq:Ekink}) at order $\lambda^{-1}$. The second term contains cross terms between $P$ and $\tilde\eta(x)$, $\tilde\pi(x)$, which arise as a price of using new variables. However, this term is ${\cal O}(\lambda)$, and thus is negligible for our purpose that evaluates the quantum correction to leading order, provided that $P$ is of order ${\cal O}(\lambda^0)$. It, however, is notable that $P^2/2E_{\rm kink}$ in this term represents the kinetic energy of a non-relativistic particle\footnote{
These terms correspond to the first two terms in the non-relativistic expansion of the kinetic energy $\sqrt{E_{\rm kink}^2+P^2}=E_{\rm kink} + P^2/2E_{\rm kink} + \cdots$. The higher order terms in the expansion manifest themselves in the higher order terms of the perturbative expansion of $\lambda$~\cite{Gervais:1975pa}. For the Lorentz symmetry of Eq.~(\ref{eq:H}), see Refs.~\cite{Gervais:1975pa,Tomboulis:1975gf}.
}. 
We also note that Eq.~(\ref{eq:H}) does not depend on $X$ explicitly as the $X$ dependence in $\tilde{H}(x-X)$ is eliminated by the $x$ integral. This fact is in accordance with the translational invariance of the theory. The third term in Eq.~(\ref{eq:H}) is independent of $X$ and $P$. $\tilde{\cal H}(x)$ is interpreted as the Hamiltonian density of the kink at $X=0$. While $\tilde{\cal H}(x)$ has a similar form as the original Hamiltonian, it is written by $\tilde\eta(x)$ and $\tilde\pi(x)$ that do not include the zero mode.

Using the new set of variables, EMT is expressed as 
\begin{align}
    T^{\mu\nu}[X,P,\tilde\eta,\tilde\pi] = T^{\mu\nu}_{\rm kink}(x-X) + \Delta\tilde{T}^{\mu\nu}[\tilde\pi(x-X),\tilde\eta(x-X)] ,
    \label{eq:Tmunu[]}
\end{align}
with
\begin{align}
\label{eq:T00eta[]}
\Delta\tilde{T}^{00}[\tilde\pi,\tilde\eta]
=&\frac{1}{2}\tilde\pi^2 
+\frac{1}{2}(\partial_x\tilde\eta)^2 
+(\partial_x\phi_{\rm kink})(\partial_x\tilde\eta)
+\lambda\phi_{\rm kink}(\phi^2_{\rm kink}-v^2)\tilde\eta 
\notag \\
&+ \frac{\lambda}{2}(3\phi^2_{\rm kink}-v^2)\tilde\eta^2
+{\cal O}(\lambda^{1/2}) ,\\
\label{eq:T11eta[]}
\Delta\tilde{T}^{11}[\tilde\pi,\tilde\eta]
=&\frac{1}{2}\tilde\pi^2 
+\frac{1}{2}(\partial_x\tilde\eta)^2 
+(\partial_x\phi_{\rm kink})(\partial_x\tilde\eta)
-\lambda\phi_{\rm kink}(\phi^2_{\rm kink}-v^2)\tilde\eta
\notag \\  
&- \frac{\lambda}{2}(3\phi^2_{\rm kink}-v^2)\tilde\eta^2
+ {\cal O}(\lambda^{1/2}) ,\\
\label{eq:T01eta[]}
\Delta\tilde{T}^{01}[\tilde\pi,\tilde\eta]=&-\tilde\pi(\partial_x\tilde\eta)   -\tilde\pi(\partial_x\phi_{\rm kink})  + {\cal O}(\lambda).
\end{align}

\subsection{Quantization}

The system described by Eq.~(\ref{eq:H}) is quantized by promoting the variables $X(t)$, $P(t)$, $\tilde\eta(x,t)$ and $\tilde\pi(t)$ to quantum operators. The Poisson brackets are promoted to the commutation relations
\begin{align}
    [X,P]=i, 
    \quad
    [\tilde\eta(x),\tilde\pi(y)]
    =i\big( \delta(x-y) - \bar\psi(x)\bar\psi(y) \big).
    \label{eq:Poisson}
\end{align}
All other commutation relations vanish. The second term in Eq.~(\ref{eq:H}) contains the cross terms between the conjugate fields. Although the order of operators has to be chosen carefully for quantizing such terms, as discussed already these terms are of order ${\cal O}(\lambda^1)$ and negligible for our purpose. 

Since the Hamiltonian~(\ref{eq:H}) does not depend on $X$ and $P$ to order that we are working, it is convenient to separate the Hilbert space $\Phi$ into the direct product as
\begin{align}
    \Phi = \Phi_X \otimes \Phi_{\tilde\eta} ,
\end{align}
where $\Phi_X$ and $\Phi_{\tilde\eta}$ represent the subspaces described by the corresponding subindices. Then, to order $\lambda^0$, the Hamiltonian is diagonalized in $\Phi_X$ and $\Phi_{\tilde\eta}$ separately. The subspace $\Phi_{\tilde\eta}$ is described by Eq.~(\ref{eq:tildeH}), and its ground state is determined without specifying the state in $\Phi_X$.

After setting the quantum state to be the ground state in $\Phi_{\tilde\eta}$, we still have arbitrariness to specify the state in $\Phi_X$. For example, one can consider eigenstates of the operator $\hat{X}$ satisfying $\hat{X}|X\rangle=X|X\rangle$, where $|X\rangle$ is assumed to be the ground state in $\Phi_{\tilde\eta}$. The matrix element of the EMT operator~(\ref{eq:Tmunu[]}) between these states is then calculated to be
\begin{align}
    \langle X | T^{\mu\nu}(x) | X' \rangle 
    = \Big( T^{\mu\nu}_{\rm kink}(x-X) +\Delta T^{\mu\nu}_{\rm kink}(x-X) \Big)
    \delta(X-X') + {\cal O}(\lambda) ,
    \label{eq:<X|T|X>}
\end{align}
with 
\begin{align}
    \Delta T^{\mu\nu}_{\rm kink}(x-X)
    = \langle X | \Delta\tilde{T}^{\mu\nu}(x) | X \rangle .
    \label{eq:DeltaT}
\end{align}
Here, $\Delta T^{\mu\nu}_{\rm kink}(x)$ is interpreted as the quantum correction of the EMT distribution around the kink at $X=0$. 

One can also consider the momentum eigenstates satisfying $\hat{P}|P\rangle=P|P\rangle$ and $\langle X | P \rangle=e^{iPX}$, where $|P\rangle$ is again assumed to be the ground state in $\Phi_{\tilde\eta}$. The matrix element of $T^{\mu\nu}(x)$ between these states is given by 
\begin{align}
    \langle P | T^{\mu\nu}(x) | P' \rangle 
    = \int dX \Big( T^{\mu\nu}_{\rm kink}(x-X) + \Delta T^{\mu\nu}_{\rm kink}(x-X) \Big)
    e^{i(P-P')X} .
    \label{eq:<P|T|P>}
\end{align}
Substituting $x=0$ into Eq.~(\ref{eq:<P|T|P>}), one sees that the Fourier transform of $T^{\mu\nu}_{\rm kink}(x) + \Delta T^{\mu\nu}_{\rm kink}(x)$ is the form factor of the kink, i.e. the GFF. In the next section, we calculate Eq.~(\ref{eq:DeltaT}). This analysis corresponds to the perturbative expansion without the zero mode. 

Further comments on the GFF are in order. Conventionally, the GFF of a spin-0 particle are defined as~\cite{Kumano:2017lhr,Hudson:2017xug,Polyakov:2018zvc}
\begin{align}
    \braket{p|T^{\mu\nu}(0)|p'}
    = \frac{K^\mu K^\nu}{K^2} \Theta_1(\Delta^2) 
    + \frac{\Delta^\mu \Delta^\nu - g^{\mu\nu} \Delta^2}{\Delta^2} \Theta_2(\Delta^2),
    \label{eq:GFF}
\end{align}
where $|p\rangle$ represents a quantum state with the Lorentz vector $p^\mu$, $K^\mu=p^\mu+p'^\mu$, $\Delta^\mu=p^\mu-p'^\mu$ and the metric tensor $g^{\mu\nu}$. Equation~(\ref{eq:GFF}) has two independent components $\Theta_1$ and $\Theta_2$. In $1+1$ dimensions, however, the projection operators satisfy $K^\mu K^\nu/K^2=g^{\mu\nu}-\Delta^\mu \Delta^\nu/\Delta^2$ and only one component does exist in the GFF, corresponding to the fact that there are no ``transverse'' directions in $1+1$ dimensions. Equation~(\ref{eq:<X|T|X>}) corresponds to the Fourier transform of this component. We also note that our analysis assumes the non-relativistic limit since it is valid only when $P$ is of order ${\cal O}(\lambda^0)$, while the kink mass~(\ref{eq:Ekink}) is of order $\lambda^{-1}$.

\section{Perturbative analysis}
\label{sec:perturbative}

\subsection{Vacuum subtraction and mass renormalization}
\begin{figure}[t]
\begin{center}
\includegraphics[scale=0.5]{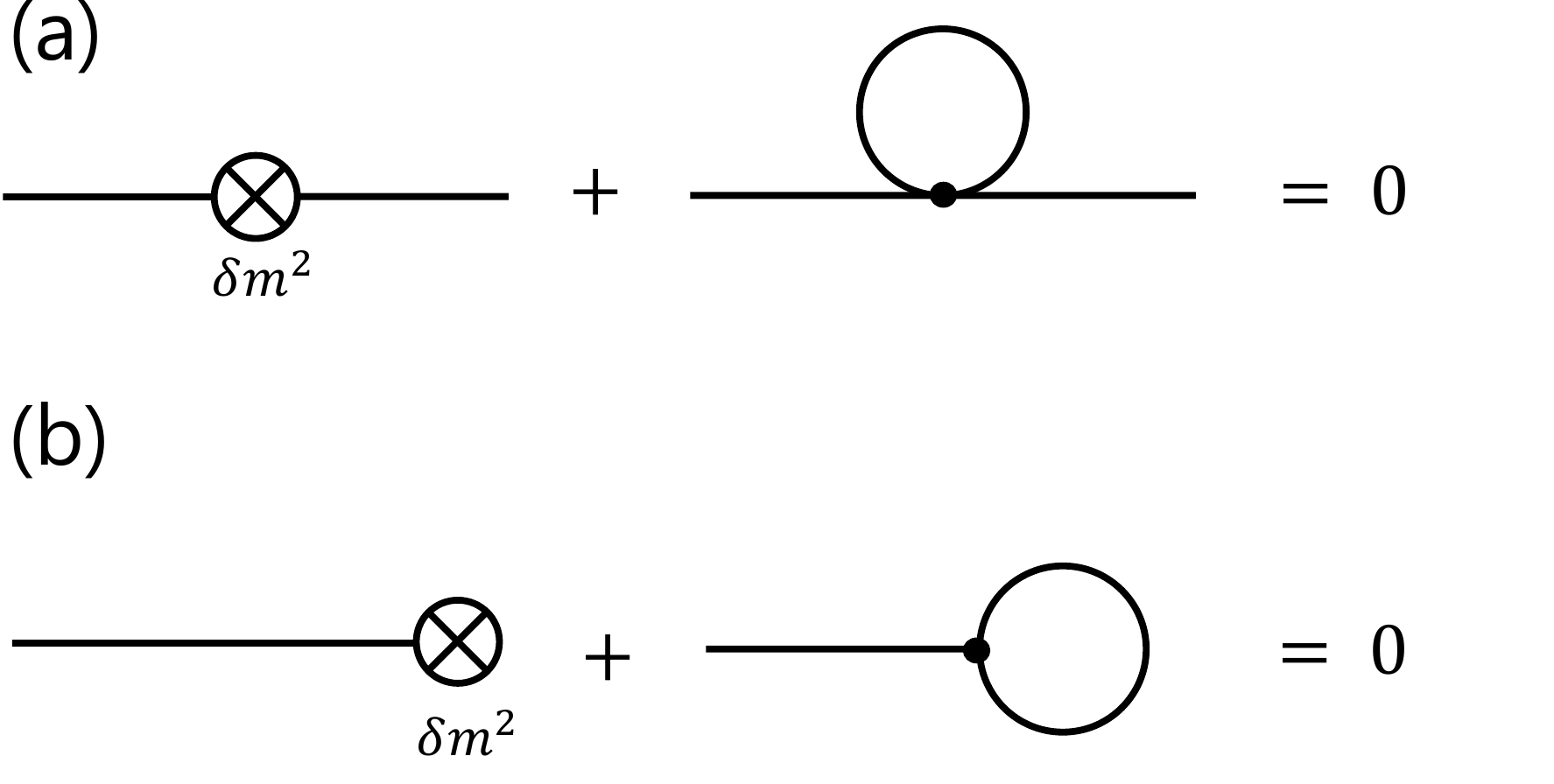}
\caption{Diagrammatic representation of the renormalization condition.}
\label{fig:massreno}
\end{center}
\end{figure}

In the analysis of Eq.~(\ref{eq:DeltaT}), we face two types of ultraviolet (UV) divergence. We remove them with the same procedure as Refs.~\cite{Dashen:1974cj,Rajaraman:1982is}\footnote{To deal with these divergences, one would first regularize the EMT operators~(\ref{eq:T00eta[]})--(\ref{eq:T01eta[]}) so that their expectation value vanishes in the vacuum sector, and then calculate their expectation values in the kink sector without the vacuum subtraction. This can be done by taking the normal ordering of Eqs.~(\ref{eq:T00eta[]})--(\ref{eq:T01eta[]}), as well as the Hamiltonian~\cite{Rajaraman:1982is,Evslin:2022opz}. It is shown that this procedure leads to the same result as that in this paper. We thank anonimous referee for notifying this point.}. We first perform the vacuum subtraction, i.e. we require that the expectation value of $T^{\mu\nu}(x)$ vanishes in the vacuum sector. This means that the expectation value in the kink sector is defined by
\begin{align}
    \langle T^{\mu\nu}(x)\rangle = \langle \tilde{T}^{\mu\nu}(x) \rangle_{\rm K} - \langle T^{\mu\nu}(x)\rangle_{\rm V},
    \label{eq:K-V}
\end{align}
where the subscripts K and V mean the expectation values for the kink and vacuum sectors, respectively, and the expectation value without a subscript is defined by Eq.~(\ref{eq:K-V}) in what follows. 

After the vacuum subtraction, Eq.~(\ref{eq:K-V}) is still UV divergent. A conventional renormalization procedure removes this divergence. It is known that the $1+1$ dimensional $\phi^4$ theory is regularized only by the mass renormalization that adds the mass counterterm
\begin{align}
\mathcal{L}_{\rm ct}=-\frac{1}{2}\delta m^2 \phi^2&=-\frac{1}{2}\delta m^2 (v^2 +2v\chi +\chi^2)
    \nonumber \\
    &=-\frac{1}{2}\delta m^2 (\phi^2_{\rm kink} +2\phi_{\rm kink}\tilde{\eta} +\tilde{\eta} ^2) ,
    \label{eq:CT}
\end{align}
to Lagrangian density\footnote{We perform the analysis in the renormalized perturbation theory, where $m$ stands for the renormalized mass. The analysis in the bare perturbation theory is discussed in App.~\ref{app:Regularization}.}. To determine $\delta m^2$ we impose the renormalization condition shown in Fig.~\ref{fig:massreno}(a), which results in 
\begin{align}
\delta m^2 &=-  \frac{3\lambda}{2L} \sum_n  \frac{1}{\sqrt{k^2_n+2m^2}},
\label{eq:deltam2}
\end{align}
where the discrete momenta $k_n$ are defined in Eq.~(\ref{eq:k_n}). This condition is equivalent to Fig.~\ref{fig:massreno}(b), i.e. vanishing of the tadpole diagram in the vacuum sector. 
We note that the common counterterm~(\ref{eq:deltam2}) is adopted to both the vacuum and kink sectors. 


As the Lagrangian density is modified by the counterterm~(\ref{eq:CT}), the EMT operator is also modified by this term. Since $\delta m^2$ is of order $\lambda^1$ as in Eq.~(\ref{eq:deltam2}), only the terms $\delta m^2 v^2$ and $\delta m^2 \phi_{\rm kink}^2$ contribute at order $\lambda^0$ in the vacuum and kink sectors, respectively. Taking this effect into account, the explicit form of $\Delta T^{\mu\nu}_{\rm kink}(x)$ is given by
\begin{align}
    \Delta T^{00}_{\rm kink}(x) &= T_1(x) + T_2(x) + T_3(x) + T_4(x),
    \label{eq:DeltaT00T}
    \\
    \Delta T^{11}_{\rm kink}(x) &= T_1(x) - T_2(x) + T_3(x) - T_4(x),  
    \label{eq:DeltaT11T}
\end{align}
with 
\begin{align}
    T_1(x) 
    =& \frac{1}{2}\langle \tilde\pi^2 \rangle_{\rm K} 
    +\frac{1}{2}\langle(\partial_1 \tilde\eta)^2\rangle_{\rm K} 
    - \frac{1}{2}\langle \tilde\pi_\chi^2 \rangle_{\rm V} 
    - \frac{1}{2}\langle(\partial_1 \chi)^2\rangle_{\rm V} ,
    \label{eq:T1}
    \\
    T_2(x)
    =& \frac{\lambda}{2}(3\phi^2_{\rm kink}-v^2)\langle\tilde\eta^2\rangle_{\rm K}
    - m^2 \langle\chi^2\rangle_{\rm V} +\frac{1}{2}\delta m^2 ( \phi^2_{\rm kink} - v^2 ) ,
    \label{eq:T2}
    \\
    T_3(x)
    =& (\partial_1 \phi_{\rm kink})\langle \partial_1 \tilde\eta \rangle_{\rm K} ,
    \label{eq:T3}
    \\
    T_4(x)
    =& \lambda\phi_{\rm kink}(\phi^2_{\rm kink}-v^2)\langle \tilde\eta  \rangle_{\rm K} .
    \label{eq:T4}
\end{align}
From time reversal symmetry, we also obtain 
\begin{align}
    \Delta T^{01}_{\rm kink}(x) &= 0.
\end{align}

In the following, we calculate Eqs.~(\ref{eq:T1})--(\ref{eq:T4}) one by one.

\subsection{$T_1(x)$ }

We start from the calculation of $T_1(x)$. The expectation values in the kink sector $\braket{\tilde\pi^2}_{\rm K}$ and $\langle(\partial_1\tilde\eta)^2\rangle_{\rm K}$ are calculated with the use of the Green function of $\tilde{H}$ given by
\begin{align}
    G(x,x';t-t') = \langle \tilde\eta(x,t) \tilde\eta(x',t') \rangle_{\rm K}
    = \int \frac{d\omega}{2\pi} \sum_{l \neq 0} e^{i\omega (t-t')}\bar\psi_l(x) \frac{i}{\omega^2-\omega^2_l+i\epsilon}\bar\psi^{*}_l(x^{\prime}),
    \label{eq:G}
\end{align}
and 
\begin{align}
    \braket{\tilde\pi(x,t) \tilde\pi(x',t') }_{\rm K} 
    =& \braket{\partial_0\tilde\eta(x,t)\partial_0\tilde\eta(x',t')}_{\rm K} \notag \\ 
    =& \partial_{t}\partial_{t'}G(x,x';t-t') 
    - \delta(t-t') \big[ \delta(x-x') - \bar\psi_0(x)\bar\psi_0(x') \big]
    \label{eq:ddGmid} \\
    =& \int \frac{d\omega}{2\pi} \sum_{l \neq 0} e^{i\omega (t-t')}\bar\psi_l(x) \frac{i\omega_l^2}{\omega^2-\omega^2_l+i\epsilon}\bar\psi^{*}_l(x') ,
    \label{eq:ddG}
\end{align}
where $\bar\psi_0(x)\bar\psi_0(x')$ in Eq.~(\ref{eq:ddGmid}) comes from the commutation relation Eq.~(\ref{eq:Poisson}) \cite{Gervais:1975pa}.

Using Eqs.~(\ref{eq:G}) and~(\ref{eq:ddG}), $\langle\tilde\pi^2\rangle_{\rm K}$  and $\langle(\partial_1\tilde\eta)^2\rangle_{\rm K}$ are calculated to be
\begin{align}
    \braket{(\partial_1 \tilde \eta)^2}_{\rm K}
    =& \lim_{x'\to x}\partial_1\partial_1' G(x,x';0)
    = \sum_{l \neq 0} \frac1{2\omega_l} |\partial_1 \bar\psi_l(x)|^2,
    \quad
    \braket{\tilde\pi^2}_{\rm K} 
    = \sum_{l \neq 0} \frac{\omega_l}2 |\bar\psi_l(x)|^2 ,
    \label{eq:<deta2>G}
\end{align}
with $\partial_1'=\partial/\partial x'$. From
\begin{align}
    |\bar\psi_q(x)|^2 
    =& \frac{|\psi_q(x)|^2}{N_q}
    \notag \\
    =& \frac1L \Big\{ 1 - \frac{3m^2}{2(q^2+2m^2)} \psi_0^2
    - \frac{3m^2}{2q^2+m^2} \psi_1^2 \Big\}\Big( 1-\frac{\delta_p'(q)}L\Big) + {\cal O}(L^{-3}),
    \\
    =& \frac1L \Big\{ 1 - \frac{3m^2}{2(q^2+2m^2)} \big( \psi_0^2 + \psi_1^2 \big) 
    - \frac{9m^4}{2(q^2+2m^2)(2q^2+m^2)} \psi_1^2 \Big\}\Big( 1-\frac{\delta_p'(q)}L\Big) 
    \notag \\
    &+ {\cal O}(L^{-3}),
    \label{eq:barpsiq2} \\
    |\partial_1 \bar\psi_q(x)|^2 
    =& \frac1L \Big\{ q^2 + 3 \big[ (\partial_1\psi_0)^2 + (\partial_1\psi_1)^2] - \frac{3m^2(\partial_1\psi_0)^2}{2(q^2+2m^2)} - \frac{3m^2(\partial_1\psi_1)^2}{2q^2+m^2}  \Big\}
    \notag \\
    &\times \Big( 1-\frac{\delta_p'(q)}L\Big)+ {\cal O}(L^{-3}),
    \label{eq:dbarpsiq2}
\end{align}
and $m^2(\psi^2_0+\psi^2_1)=2((\partial_1\psi_0)^2 + (\partial_1\psi_1)^2)$, one obtains 
\begin{align}
    \braket{\tilde\pi^2}_{\rm K}  + \braket{(\partial_1 \tilde \eta)^2}_{\rm K}
    =& \frac{1}{2L}\sum_{n} \Big( \sqrt{q^2_n+2m^2} + \frac{q_n^2}{\sqrt{q_n^2+2m^2}}\Big) \Big( 1-\frac{\delta_p'(q_n)}L\Big)
    \notag \\
    & -\frac32 D_1(\partial_1 \psi_0)^2
    +\Big(\frac{\sqrt{3}}{4}-\frac32 D_2\Big)\Big( \frac{3m^2}2\psi_1^2 + (\partial_1 \psi_1)^2\Big),
    \label{eq:<deta2>}
\end{align}
with
\begin{align}
    D_1 =& \frac{1}{L} \sum_{n} \frac{m^2}{2\sqrt{q^2_n+2m^2}(q^2_n+2m^2)} \Big( 1-\frac{\delta_p'(q_n)}L\Big)
    \nonumber \\
    &\xrightarrow[L\to\infty]{} \int_{-\infty}^\infty \frac{dq}{2\pi} \frac{m^2}{2\sqrt{q^2+2m^2}(q^2+2m^2)} = \frac{1}{4\pi},
    \label{eq:D1}
    \\
    D_2 =& \frac{1}{L}\sum_{n} \frac{m^2 }{\sqrt{q^2_n+2m^2}{(2q^2_n+m^2)}} \Big( 1-\frac{\delta_p'(q_n)}L\Big)
    \nonumber \\
    &\xrightarrow[L\to\infty]{} \int_{-\infty}^\infty \frac{dq}{2\pi} \frac{m^2}{\sqrt{q^2+2m^2}(2q^2+m^2)} = \frac{\sqrt{3}}{9} .
    \label{eq:D2}
\end{align}
Here, the sums in Eqs.~(\ref{eq:D1}) and~(\ref{eq:D2}) are convergent and they are replaced with the integrals in the $L\to\infty$ limit. 

The first term in Eq.~(\ref{eq:<deta2>}) is UV divergent. The divergence is removed by the subtraction of the vacuum sector given by
\begin{align}
    \braket{\tilde\pi_\chi^2}_{\rm V} + \braket{(\partial_1 \chi)^2}_{\rm V}  
    = \frac{1}{2L} \sum_{n} \Big( \sqrt{k^2_n+2m^2} + \frac{k^2_n}{\sqrt{k^2_n+2m^2}} \Big) ,
\end{align}
As shown in App.~\ref{sec:MNC}, the results of the subtraction in the MNC are given by
\begin{align}
    &\frac{1}{L}\sum_{n} \sqrt{q^2_n+2m^2} \Big( 1-\frac{\delta_p'(q_n)}L\Big)-\frac{1}{L}\sum_{n}\sqrt{k^2_n+2m^2} 
    \xrightarrow[L\to\infty]{} -\frac{3\sqrt{2}m}{\pi L} ,
    \label{eq:V1}
     \\
    &\frac{1}{L}
    \sum_{n} \frac{q_n^2}{\sqrt{q^2_n+2m^2}} \Big( 1-\frac{\delta_p'(q_n)}L\Big)-\frac{1}{L}\sum_{n}\frac{k_n^2}{\sqrt{k^2_n+2m^2}}
    \xrightarrow[L\to\infty]{} -\frac{3\sqrt{2}m}{\pi L} ,
    \label{eq:V3}
     \\
    &\frac{1}{L}\sum_{n} \frac1{\sqrt{q^2_n+2m^2}} \Big( 1-\frac{\delta_p'(q_n)}L\Big)-\frac{1}{L}\sum_{n}\frac1{\sqrt{k^2_n+2m^2}} \xrightarrow[L\to\infty]{} 0 . 
    \label{eq:V2}
\end{align}

Accumulating these results, one obtains 
\begin{align}
    T_1(x) 
    =&
    -\frac{3\sqrt{2}m}{2 \pi L}
    -\frac{3}{4}D_1(\partial_1 \psi_0)^2
    +\frac{1}{2}\Big(\frac{\sqrt{3}}{4}-\frac{3}{2}D_2\Big) \Big( \frac{3m^2}2 \psi_1^2 + (\partial_1 \psi_1)^2 \Big)
    \notag \\
    =& -\frac{3\sqrt{2}m}{2\pi L} 
    + \frac A2 (\partial_1 \psi_0)^2 + \frac B2 \Big( \frac{3m^2}2 \psi_1^2 + (\partial_1 \psi_1)^2 \Big)   \notag \\
    =& -\frac{3\sqrt{2}m}{2\pi L}  +Bm^2 {\rm sech}^2 \frac{mx}{\sqrt{2}} +\Big(A-\frac74 B \Big)m^2 {\rm sech}^4 \frac{mx}{\sqrt{2}}
    +(-A+B) m^2 {\rm sech}^6 \frac{mx}{\sqrt{2}} ,
\end{align}
with 
\begin{align}
A &= -\frac{3}{2}D_1=-\frac{3}{8\pi}, 
\label{eq:A}\\
B &=\frac{\sqrt{3}}{4}-\frac{3}{2}D_2=\frac{\sqrt{3}}{12}. 
\label{eq:B}
\end{align}

\subsection{$T_2(x)$}

Next, let us calculate $T_2(x)$. $\langle\tilde\eta^2(x)\rangle_{\rm K}$ and $\langle\chi^2(x)\rangle_{\rm V}$ are calculated to be
\begin{align}
    \langle \tilde\eta(x)^2 \rangle_{\rm K}
    =& \lim_{x'\to x} G(x,x';0) 
    = \sum_{l \neq 0} \frac{1}{2\omega_l} |\bar\psi_l(x)|^2  \notag
    \\
    =&  \frac{1}{2L}\sum_{n}\frac{1}{\sqrt{q^2_n+2m^2}} \Big( 1-\frac{\delta_p'(q_n)}L\Big)
    +A\psi^2_0 + B\psi^2_1 ,
    \label{eq:<eta2>}
    \\
    \braket{\chi^2(x)}_{\rm V} 
    =& \frac{1}{2L} \sum_{n} \frac{1}{\sqrt{k^2_n+2m^2}}
    = -\frac1{3\lambda} \delta m^2 . 
\label{eq:<chi2>}
\end{align}
Substituting them into $T_2(x)$, we obtain
\begin{align}
    T_2(x) 
    =& \frac{\lambda}{2}(3\phi^2_{\rm kink}-v^2) ( \braket{ \tilde\eta^2}_{\rm K} - \braket{\chi^2}_{\rm V} ) 
    \notag \\
    =& 
    \frac{m^2}{2} (3\bar\phi^2_{\rm kink}-1) ( A \psi^2_0 + B \psi^2_1 )
    \notag \\
    =& B  m^2 {\rm sech}^2 \frac{mx}{\sqrt{2}}
    + \Big(A- \frac52 B \Big)  m^2 {\rm sech}^4 \frac{mx}{\sqrt{2}}
    - \frac{3}{2} (A- B)  m^2 {\rm sech}^6 \frac{mx}{\sqrt{2}},
\end{align}
where $\bar\phi_{\rm kink}(y)=(\sqrt{\lambda} / m) \phi_{\rm kink}(y;0) $ and we used Eq.~(\ref{eq:V2}) for the vacuum subtraction.

\subsection{$T_3(x)$ and $T_4(x)$}
\label{sec:tad}

\begin{figure}[t]
\begin{center}
\includegraphics[scale=0.5]{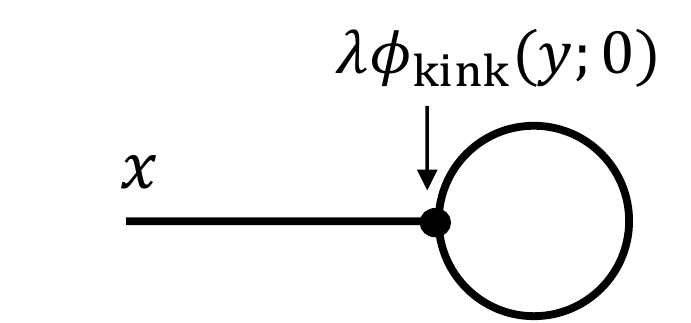}
\caption{Tadpole diagram contributing to $\braket{\tilde\eta(x)}_{\rm K}$.}
 \label{fig:tadpole}
\end{center}
\end{figure}

Finally, we calculate the terms including one-point correlation function $T_3(x)$ and $T_4(x)$. The coefficients of these terms are of order $\lambda^{-1/2}$. They have order $\lambda^0$ contribution through $\braket{\tilde\eta(x)}_{\rm K}$ obtained from the tadpole diagram in Fig.~\ref{fig:tadpole}, where the three-point interaction $\lambda\phi_{\rm kink}$ is of order $\lambda^{1/2}$. The one-point function in the vacuum sector $\braket{\chi(x)}_{\rm V}$ vanishes by mass renormalization. The expectation value of $\tilde\eta(x)$ is calculated to be
\begin{align}
    \braket{\tilde\eta(x)}_{\rm K}
    =&-i\int dy^2 
    \phi_{\rm kink}(y;0) \big \{ 3\lambda \braket{\tilde\eta(x)\tilde\eta(y)}_{\rm K} \braket{\tilde\eta(y)\tilde\eta(y)}_{\rm K} +\delta m^2 \braket{\tilde\eta(x)\tilde\eta(y)}_{\rm K}  \big \} 
    \notag \\
    =&
    -i \int dy \phi_{\rm kink}(y;0) \tilde{G}(x,y) \{ 3\lambda  {G}(y,y) + \delta m^2 \}
    \notag\\
    =
    &-3i\lambda\int dy\phi_{\rm kink} (y;0)\tilde{G}(x,y) \{A\psi^2_0+B\psi^2_1 \} ,
    \label{eq:<eta>}
\end{align}
where $\tilde{G}(x,y)=\int dt' G(x,y;t-t')$. In the last equality, we used Eqs.~(\ref{eq:deltam2}), (\ref{eq:V2}) and~(\ref{eq:<eta2>}). 
Using $\psi_0^2(x)=1-2\bar\phi^{2}_{\rm kink}(x)+\bar\phi^{4}_{\rm kink}(x)$ and $\psi_1^2(x)=\bar\phi^{2}_{\rm kink}(x)-\bar\phi^{4}_{\rm kink}(x)$, Eq.~(\ref{eq:<eta>}) is further rewritten as
\begin{align}
    \braket{\tilde\eta(x)}_{\rm K}
    =- 3im\sqrt{\lambda} \big\{ A\big(H_1(x)-2H_3(x)+H_5(x)\big) + B\big(H_3(x)-H_5(x)\big) \big\},
    \label{eq:<eta>=H}
\end{align}
with 
\begin{align}
    H_i(x) = \int dy 
    \bar\phi^{i}_{\rm kink}(y) \tilde{G}(x,y).
    \label{eq:H_i}
\end{align}

Using analytic forms of Eq.~(\ref{eq:H_i}) given in App.~\ref{app:tadpole}, one obtains
\begin{align}
\label{eq:<eta>=phi}
    \braket{\tilde\eta(x)}_{\rm K} 
    =& 
    -\frac{\sqrt{\lambda}}{m} \Big\{ (A-B) \bar\phi_{\rm kink} ( 1-\bar\phi_{\rm kink}^2) + \frac32 B x \partial_1 \bar\phi_{\rm kink} \Big\} .
\end{align}
This result gives  
\begin{align}  
    T_3(x) =&
  -( A-B ) \frac{m^2}2 \psi_0^2 ( 1-3\bar{\phi}_{\rm kink}^2)
    -\frac32 B (\partial_1\bar{\phi}_{\rm kink})\partial_1( x \partial_1\bar{\phi}_{\rm kink}) \notag \\
    =& \Big(A-\frac74 B \Big) m^2 {\rm sech}^4 \frac{mx}{\sqrt{2}} -\frac{3}{2}(A-B) m^2 {\rm sech}^6 \frac{mx}{\sqrt{2}} \notag \\
    &+\frac{3\sqrt{2}}{4} B m^3 x \tanh{\frac{mx}{\sqrt{2}}}  {\rm sech}^4 \frac{mx}{\sqrt{2}} ,\\
    T_4(x)
    =& 
    \frac12 ( A-B ) (\partial_1\psi_0)^2 
    +\frac32 B m^2\bar{\phi}_{\rm kink} (1-\bar{\phi}_{\rm kink}^2) x \partial_1\bar{\phi}_{\rm kink} \notag \\
    =& ( A-B )  m^2 {\rm sech}^4 \frac{mx}{\sqrt{2}} - (A-B) m^2 {\rm sech}^6 \frac{mx}{\sqrt{2}} \notag \\
    &+\frac{3\sqrt{2}}{4} B m^3 x \tanh{\frac{mx}{\sqrt{2}}}  {\rm sech}^4 \frac{mx}{\sqrt{2}}.
\end{align}
As discussed in App.~\ref{app:tadpole}, in the analysis of Eq.~(\ref{eq:H_i}) there appear surface terms from partial integrals. Equation~(\ref{eq:<eta>=phi}) is the result obtained neglecting these terms. We note that the vanishing of the surface terms is most clearly justified with the APBC, although it would be valid for any case.

Equation~(\ref{eq:<eta>=phi}) also allows us to calculate the quantum correction to the topological charge density~(\ref{eq:topol}) as discussed in App.~\ref{sec:topological}.

\section{Result}
\label{sec:result}

\begin{figure}[t]
\begin{center}
\includegraphics[scale=0.4]{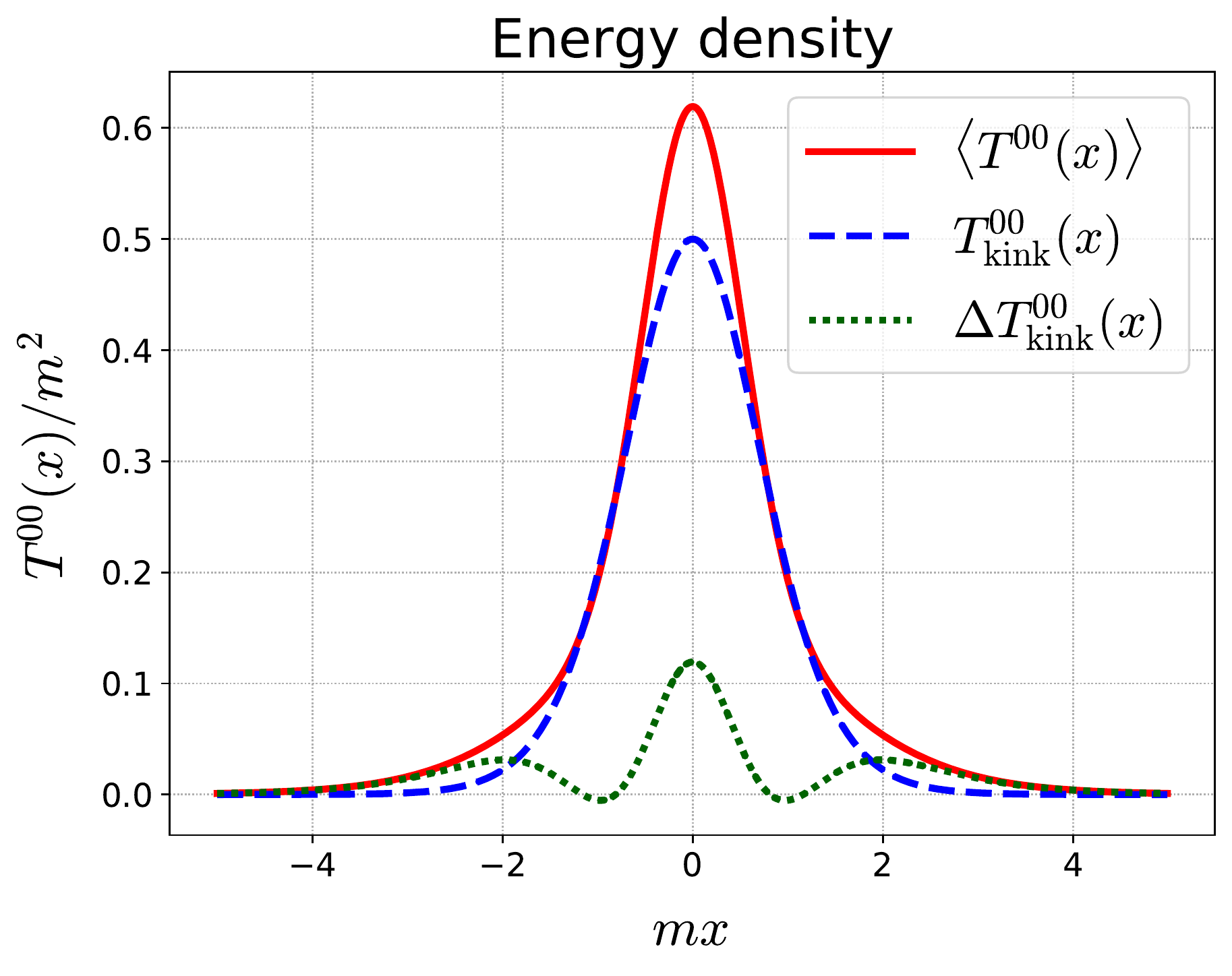}
\caption{Energy density around the kink $\braket{T^{00}(x) }$ at $\lambda/m^2=1$. The classical value $T^{00}_{\rm kink}(x)$ and the quantum correction $\Delta T^{00}_{\rm kink}(x)$ in the $L\to\infty$ limit are also shown by the dashed and dotted lines, respectively. }
 \label{fig:T00}
\end{center}
\end{figure}
Accumulating these results, the expectation value of EMT to one-loop order is obtained as
\begin{align}
    \braket{T^{\mu\nu}(x) } =&T^{\mu\nu}_{\rm kink}(x)+\Delta T^{\mu\nu}_{\rm kink}(x),
    \label{eq:tildeT00final}
    \\
    \Delta T^{00}_{\rm kink}(x)=& \frac{\sqrt{3}}{6} m^2 {\rm sech}^2 \frac{mx}{\sqrt{2}} - \Big (\frac{3}{2 \pi } +\frac{7\sqrt{3}}{12} \Big )m^2 {\rm sech}^4 \frac{mx}{\sqrt{2}}
    \notag\\
    &+5 \Big ( \frac{3}{8 \pi }+ \frac{\sqrt{3}}{12} \Big )m^2 {\rm sech}^6 \frac{mx}{\sqrt{2}}  +\frac{\sqrt{6}}{8}  m^3 x \tanh{\frac{mx}{\sqrt{2}}}  {\rm sech}^4 \frac{mx}{\sqrt{2}} 
    \notag \\
    &-\frac{3\sqrt{2}m}{2\pi L},
    \label{eq:T00final}
    \\
    \Delta T^{11}_{\rm kink}(x) 
    =& -\frac{3\sqrt{2}m}{2\pi L} .
    \label{eq:T11final} 
\end{align}
In Fig.~\ref{fig:T00}, we show the behavior of Eqs.~(\ref{eq:tildeT00final}) and (\ref{eq:T00final}) in the $L\to\infty$ limit together with the classical value $T^{00}_{\rm kink}(x)$ at $\lambda/m^2=1$.
By taking the spatial integral of $\braket{T^{00}(x) }$, we obtain the total energy 
\begin{align}
    \int_{-L/2}^{L/2} dx \braket{T^{00}(x) } &= E_{\rm kink} +m \Big ( \frac{\sqrt{6}}{12}  -\frac{3\sqrt{2}}{2\pi}    \Big ) ,
    \label{eq:E}
\end{align}
which reproduces the result in Ref.~\cite{Dashen:1974cj}.

A notable feature of Eqs.~(\ref{eq:tildeT00final})--(\ref{eq:T11final}) is that all $x$ dependence cancels out in Eq.~(\ref{eq:T11final}) and $\braket{T^{11}(x)}$ becomes a constant. This result is in agreement with the momentum conservation in static systems, $\partial_1 T^{11}(x)=0$, which is also interpreted as the equilibration of the force. While the energy density, i.e. Eq.~(\ref{eq:T00final}), has been investigated in the same model in Refs.~\cite{Goldhaber:2001rp,Martin:2022pri}, $\braket{T^{11}(x)}$ is not analyzed there. Our result~(\ref{eq:T00final}) does not agree with any of them\footnote{In Ref.~\cite{Goldhaber:2001rp,Martin:2022pri}, the definition of $m$ is different from ours. Their results are comparable to ours with a replacement $m^2\to2m^2$.}. The calculation of $\braket{T^{11}(x)}$ and a confirmation of the momentum conservation would be used for a check of the validity of each analysis. Although the reproduction of their analyses is difficult, we give some arguments in App.~\ref{sec:LMR}. 

Although Eq.~(\ref{eq:T11final}) is obtained to leading order in $1/L$, it is easily confirmed that $\partial_1 \braket{T^{11}(x)}=0$ holds even to higher orders in $1/L$ as follows. The higher order terms in $1/L$ come from Eqs.~(\ref{eq:barpsiq2}) and~(\ref{eq:dbarpsiq2}), and also Eqs.~(\ref{eq:V1}) and~(\ref{eq:V2}). Among their effects on the final result, the modifications of $A$ and $B$ in Eqs.~(\ref{eq:A}) and~(\ref{eq:B}) do not affect the cancellation of each term in $\Delta T^{11}_{\rm kink}(x)$. Also, Eq.~(\ref{eq:V2}) becomes nonzero at order $1/L^2$, and it modifies $T_2(x)$, $T_3(x)$ and $T_4(x)$. However, one can easily verify that this effect cancels out in $x$-dependent terms. Therefore, $\braket{T^{11}(x)}$ is a constant even to higher order in $1/L$, as it should be from the momentum conservation.

Equation~(\ref{eq:T11final}), however, is nonzero. Equation~(\ref{eq:T00final}) also contains the same term proportional to $1/L$. The term in Eq.~(\ref{eq:T00final}) contributes to the total energy Eq.~(\ref{eq:E}) and is mandatory to reproduce the result of Ref.~\cite{Dashen:1974cj}. However, this term vanishes if one defines the energy density of the kink as the $L\to\infty$ limit of Eq.~(\ref{eq:T00final}). The total energy defined from this energy density  
\begin{align}
    \int_{-\infty}^{\infty}dx\lim_{L\to\infty}\braket{T^{00}(x)}
    = E_{\rm kink} + \frac{\sqrt{6}}{12} m,
    \label{eq:Einf}
\end{align}
thus contradicts the one of Ref.~\cite{Dashen:1974cj}. This result raises the question of what is the correct total energy of the kink. We note that the quantum correction in Ref.~\cite{Dashen:1974cj} is negative, while Eq.~(\ref{eq:Einf}) is positive.

There is another issue concerning the $1/L$ term. Since the stress tensor $\braket{T^{11}(x)}$ represents the force acting on each space point, this force does the work when the system size $L$ is varied. More specifically, by varying the system size from $L$ to $L+\Delta L$, the total energy of the system should be reduced by $T^{11}\Delta L$ due to the work. As a result, the total energy must have a term proportional to $\int_L dL' T^{11}(L')$~
\footnote{For the case of Casimir effect, this correspondence between the stress and internal energy is fullfilled~\cite{Brown:1969na,Kitazawa:2019otp}.}. Substiting $T^{11}(L)\sim 1/L$ into this term leads to its logarithmic divergence in the $L\to\infty$ limit, which, however, constradicts the existence of the finite total energy in this limit. There are several possibilities to explain this contradiction. One of them is that $T^{11}(x)$ would not be interpreted as the force in this system, or the system investigated here would not physically correspond to the one in which kinks and anti-kinks are aligned alternately with the interval $L$. It would also be possible that the regularization based on the MNC with the CCM causes problems in the vacuum subtraction~\cite{Rebhan:1997iv,Evslin:2022opz}. We also note that our analysis relies on the nonrelativistic approximation since it is justified only when $P$ is of order ${\cal O}(\lambda^0)$ as discussed in Sec.~\ref{sec:CCM}. The clarification of the problem, however, is beyond our present understanding, and we leave it for future study.

\section{Summary and outlook}
\label{sec:discussions}

In this study, we have explored the EMT distribution around the kink in the $1+1$ dimensional $\phi^4$ theory to one-loop order. Our final result is given in Eqs.~(\ref{eq:T00final}) and (\ref{eq:T11final}). This result is consistent with the momentum conservation $\partial_1\braket{\tilde{T}^{11}(x)}=0$. The spatial integral of $\braket{\tilde{T}^{00}(x)}$ reproduces the total energy in Ref.~\cite{Dashen:1974cj}, while our result obtained in a finite system of length $L$ contains a constant term proportional to $1/L$, whose physical interpretation deserves further investigation.

There are many future extensions of the present study. Investigations of the kink and localized structures in other $1+1$ dimensional models are straightforward ones among them. An example is the sine-Gordon model, which has a stable kink solution and the time-dependent solution called the breather mode~\cite{Dashen:1974cj}. While the quantum correction to their total energy has been investigated~\cite{Dashen:1975hd}, EMT distribution has not been analyzed so far. Their analysis will be reported in the forthcoming publication~\cite{Ito:prep}. Next, exploring the quantum effects on the localized structures in higher-dimensional systems is a further interesting subject. For example, the $2+1$ dimensional $\phi^4$ theory has a classical solution
\begin{align}
    \phi(x,y) = \phi_{\rm kink}(x;X),
    \label{eq:2dim}
\end{align}
having the translational invariance in the $y$ direction, which is the surface connecting two vacua. By quantizing this system, the position of the kink is obscured. Although this effect leads to instability for an infinitely-long surface, the problem will be well defined when the positions of the surface are fixed by hand at two points, say $y=\pm R/2$. An investigation of the EMT distribution in this system will give us novel insights into the quantum effects on the surface. The problem would also be extended to a $3+1$ dimensional system, where the classical solutions can have string-like structures, such as the vortex solution in the Abelian-Higgs model~\cite{Yanagihara:2019foh}. The analysis of quantum effects in this system will provide us with a microscopic basis of the effective string models~\cite{Luscher:1980ac,Luscher:2004ib,Polchinski:1991ax,Kuti:2005xg}, as well as the numerical results of flux tube~\cite{Kuti:2005xg,Cardoso:2013lla,Yanagihara:2018qqg}.

\acknowledgments
The authors thank Jarah Evslin and Shunzo Kumano for valuable comments.
They are also grateful to Teiji Kunihiro and Hiroshi Suzuki for their encouragement. This work was supported by JST SPRING, Grant Number JPMJSP2138, and JSPS KAKENHI (Grants No.~JP19H05598, No.~20H01903, No.~22K03619).
\appendix

\section{Topological charge density}
\label{sec:topological}

\begin{figure}[t]
\begin{center}
\includegraphics[scale=0.4]{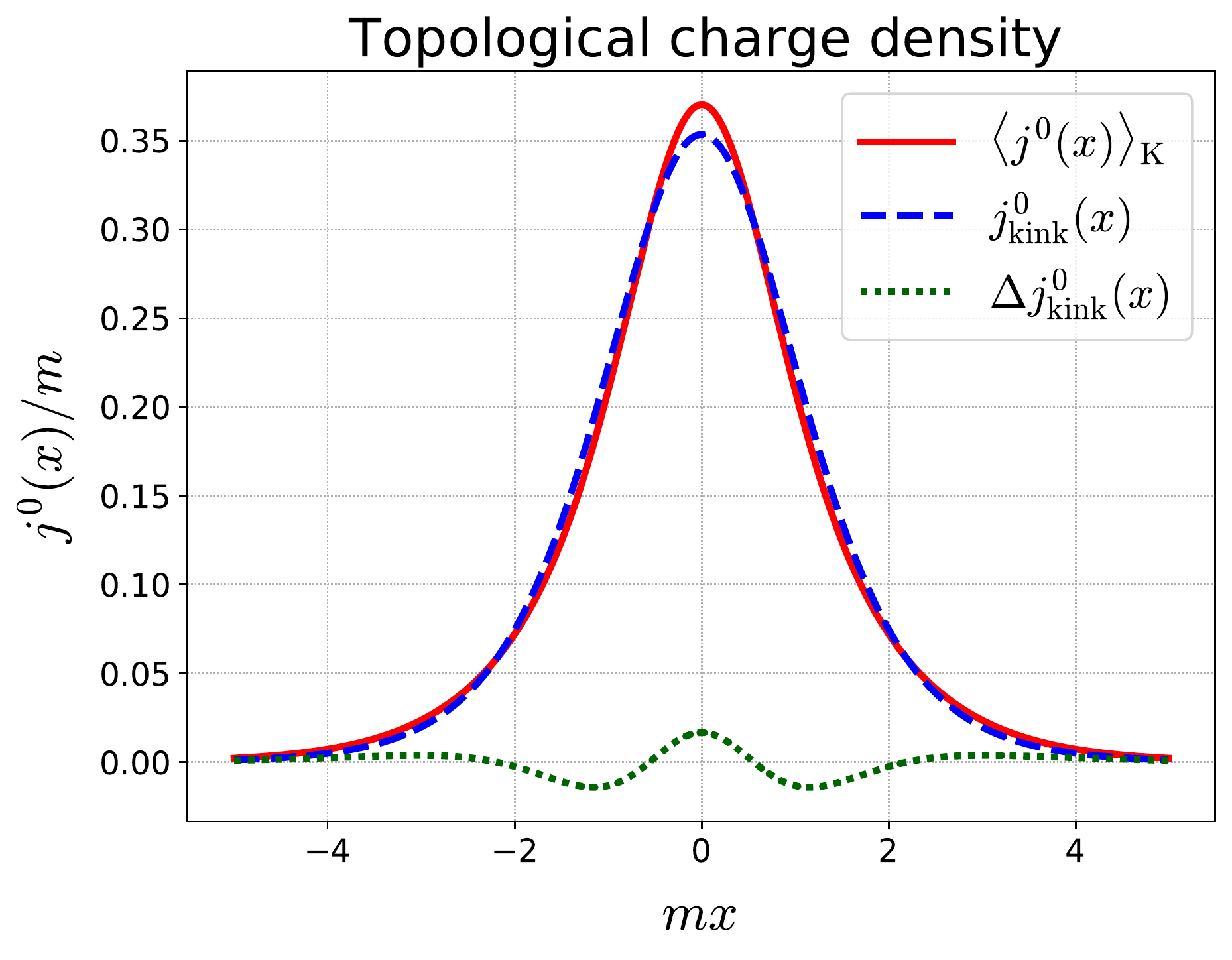}
\caption{Topological charge density around a kink at $\lambda/m^2=1$.}
 \label{fig:j0}
\end{center}
\end{figure}
In this appendix, we calculate the 1-loop correction to the topological charge density~(\ref{eq:topol}) in the kink sector.
From Eq.~(\ref{eq:<eta>}), the expectation value of $j^0(x)$ in kink sector is given by
\begin{align}
\label{eq:j0final}
\braket{j^0(x)}_{\rm K} =&  j^{0}_{\rm kink }(x)+ \Delta j^{0}_{\rm kink }(x) ,
\\
\Delta j^{0}_{\rm kink }(x) =& \frac{1}{2v} \langle \partial_1 \tilde{\eta} \rangle_{\rm K} \notag \\
=&\frac{\lambda}{2m} \big [ (\sqrt{2}A-\frac{7\sqrt{2}}{4}B ) {\rm sech}^2 \frac{mx}{\sqrt{2}} +\frac{3\sqrt{2}}{2}(-A+B) {\rm sech}^4 \frac{mx}{\sqrt{2}} \notag \\
&+\frac{3}{2}B m x \tanh{\frac{mx}{\sqrt{2}}} {\rm sech}^2 \frac{mx}{\sqrt{2}}  \big ].
\label{eq:j0final2}
\end{align}
The behavior of Eqs.~(\ref{eq:j0final}) and~(\ref{eq:j0final2}) at $\lambda/m^2=1$ is shown in Fig.~\ref{fig:j0}. The spatial integral of Eq.~(\ref{eq:j0final2}) is given by 
\begin{align}
\int dx \Delta j^{0}_{\rm kink }(x)  =0.
\end{align}
which leads to $\int dx \braket{j^0(x)}_{\rm K} =1$.

\section{Mass renormalization}
\label{app:Regularization}

In this Appendix, we discuss the dependence of our results on the mass renormalization condition and clarify their mutual relation. It is also shown that the bare perturbation theory gives the same result as Eqs.~(\ref{eq:tildeT00final})--(\ref{eq:T11final}).

To resolve these issues, we first point out that the terms arising from the mass counterterm cancel out in $\braket{T^{11}(x)}$. This can be checked by formally accumulating terms including $\delta m^2$ in Eq.~(\ref{eq:DeltaT11T}). Such terms exist in $T_2(x)$, $T_3(x)$ and $T_4(x)$. First, $T_2(x)$ contains
\begin{align}
   \delta T_2(x) = \frac{1}{2}\delta m^2 ( \phi^2_{\rm kink} - v^2 ) 
   = - \frac{m^2}{2\lambda} \delta m^2 {\rm sech}^2 \frac{mx}{\sqrt{2}}.
   \label{eq:deltaT2}
\end{align}
Next, in $T_3(x)$ and $T_4(x)$, such terms arise from the first or second line of Eq.~(\ref{eq:<eta>})
\begin{align}
    \braket{\delta \tilde\eta(x)}_{\rm K}
    = -i \delta m^2 \int \frac{dy}{2\pi} \phi_{\rm kink}(y;0) \tilde{G}(x,y) 
    = -i \frac{m}{\sqrt{\lambda}} \delta m^2 H_1(x) ,
    \label{eq:deltaeta}
\end{align}
that gives
\begin{align}
    \delta T_3(x) =& 
    (\partial_1 \phi_{\rm kink})\langle \partial_1 \delta\tilde\eta \rangle_{\rm K} 
    = \frac{m^2 \delta m^2}{2\lambda}\Big ( - {\rm sech}^4 \frac{mx}{\sqrt{2}}+\frac{mx}{\sqrt{2}}\tanh{\frac{mx}{\sqrt{2}}}  {\rm sech}^4 \frac{mx}{\sqrt{2}} \Big ) ,    
    \label{eq:deltaT3}
    \\
    \delta T_4(x) =& 
    \lambda\phi_{\rm kink}(\phi^2_{\rm kink}-v^2)\langle \delta \tilde\eta \rangle_{\rm K} 
    \notag \\
    =& \frac{m^2 \delta m^2}{2\lambda}\Big (  {\rm sech}^2 \frac{mx}{\sqrt{2}} - {\rm sech}^4 \frac{mx}{\sqrt{2}}+\frac{mx}{\sqrt{2}}\tanh{\frac{mx}{\sqrt{2}}}  {\rm sech}^4 \frac{mx}{\sqrt{2}} \Big ) .
    \label{eq:deltaT4}
\end{align}
From Eqs.~(\ref{eq:deltaT2}), (\ref{eq:deltaT3}) and (\ref{eq:deltaT4}) one finds 
\begin{align}
    - \delta T_2(x) + \delta T_3(x) - \delta T_4(x) = 0 .
    \label{eq:deltaT234}
\end{align}
which means that these terms cancel out in Eq.~(\ref{eq:DeltaT11T}).

From Eq.~(\ref{eq:deltaT234}), it is concluded that $\braket{T^{11}(x)}$ does not depend on $\delta m^2$, and hence the renormalization condition. In particular, the momentum conservation $\partial_1 \braket{T^{11}(x)}=0$ is always satisfied. Equation~(\ref{eq:deltaT234}) also tells us that Eq.~(\ref{eq:T11final}) is obtained in the bare perturbation theory (BPT), where the mass counterterm does not exist.

Next, let us focus on $\braket{T^{00}(x)}$. In this case, the change of the renormalized mass modifies the classical term $T_{\rm kink}^{00}(x)$, which gives rise to additional terms at order $\lambda^0$. To clarify the discussion, let us consider two renormalization conditions that give different renormalized masses $m_1$ and $m_2$, whose difference $m_1-m_2$ is of order $\lambda$. Then, in each renormalization, the classical energy density is given by $m_1^4/(2\lambda) \mbox{sech}^4 (m_1x/\sqrt2)$ and $m_2^4/(2\lambda) \mbox{sech}^4 (m_2x/\sqrt2)$, respectively, whose difference is
\begin{align}
    (m_1-m_2) \frac{\partial T_{\rm kink}^{00}(x) }{\partial m} ,
    \label{eq:dTdm}
\end{align}
where the value of mass in $\partial T_{\rm kink}^{00}(x) /\partial m$ is irrelevant at order $\lambda^0$.
On the other hand, the value of mass counterterms $\delta m_1^2$ and $\delta m_2^2$ in each renormalization differ by $\delta m_1^2-\delta m_2^2 = m_1^2-m_2^2 \sim (m_1-m_2)m_1$. The difference of Eq.~(\ref{eq:DeltaT00T}) coming from it is $\delta T_2(x)+\delta T_3(x)+\delta T_4(x)$ but $\delta m^2$ is replaced with $\delta m_1^2-\delta m_2^2$. It is shown that this modification exactly cancels out with Eq.~(\ref{eq:dTdm}). This can be shown from the relation
\begin{align}
    -\frac{\delta m^2}{2m} \frac{\partial T_{\rm kink}^{00}(x) }{\partial m}
    = \delta T_2(x) + \delta T_3(x) + \delta T_4(x) + {\cal O}(\lambda),
    \label{eq:dTdm=T}
\end{align}
that is obtained by an explicit calculation of the left-hand side.

Equation~(\ref{eq:dTdm=T}) also tells us that Eq.~(\ref{eq:T00final}) is obtained even in the BPT. In the BPT, we use the bare mass $m_0$ that is related to the renormalized mass $m$ as $m^2=m_0^2+\delta m^2$, while the mass counterterm is not introduced. In this case, since the mass in $T_{\rm kink}^{00}(x)$ is $m_0$, Eq.~(\ref{eq:dTdm=T}) appears at order $\lambda^0$ when $T_{\rm kink}^{00}(x)$ is rewritten by $m$, which is exactly the term coming from the mass counterterm in the renormalized perturbation theory. Therefore, the results in the bare and renormalized perturbation theories are the same at order $\lambda^0$ as they should be.

\section{Mode-number cutoff}
\label{sec:MNC}

\begin{figure}[t]
\begin{center}
\includegraphics[scale=0.45]{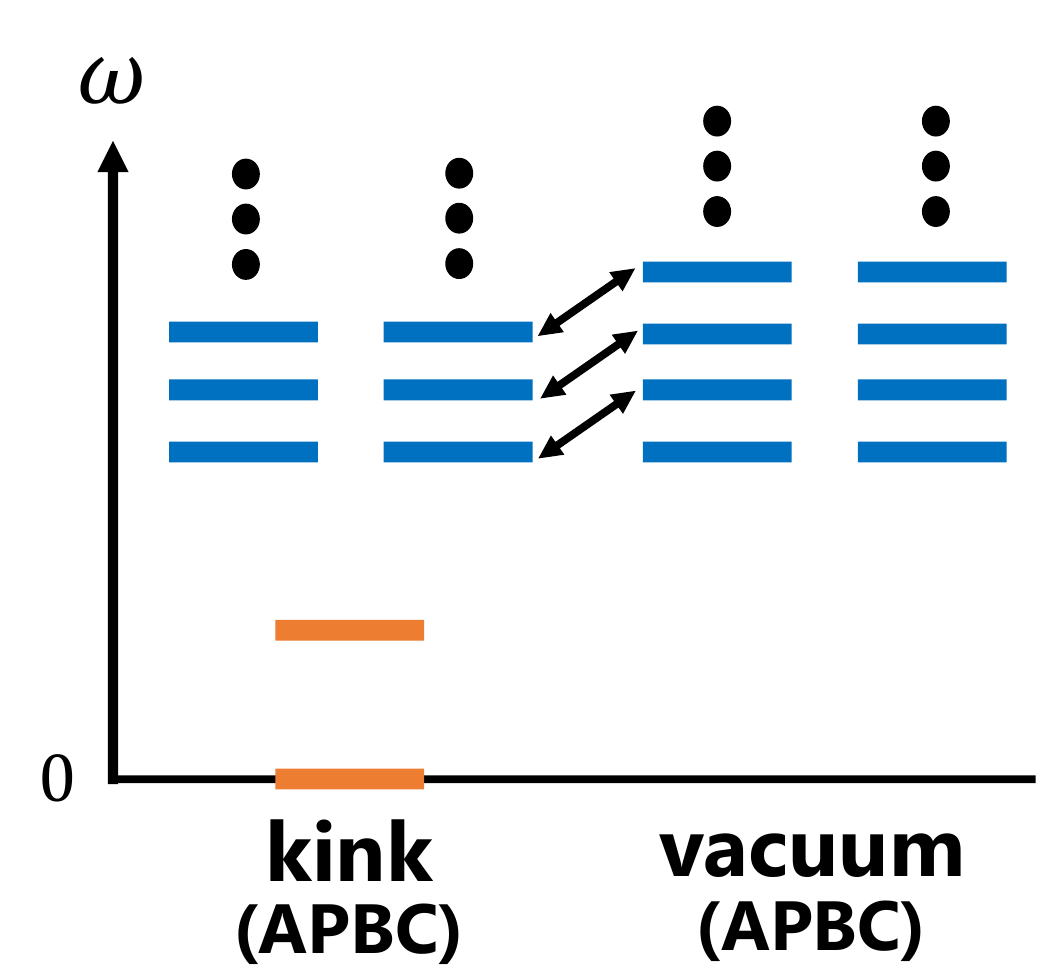}
\caption{Eigenvalue distributions in the kink and vacuum sectors for the APBC. The kink sector has two bound states shown by the orange lines. The continuum spectra shown by the blue lines are doubly degenerated. On the right-hand side of Eq.~(\ref{eq:MNC3}), the subtraction is taken between the modes connected by the arrows. } 
\label{fig:MNC}
\end{center}
\end{figure}

In this appendix we derive Eqs.~(\ref{eq:V1}) and~(\ref{eq:V2}). For this, we use the mode-number cutoff (MNC) prescription. We refer to Refs.~\cite{Dashen:1974cj,Rajaraman:1982is,Rebhan:1997iv} for a more detailed discussion on the MNC. In particular, see Ref.~\cite{Rebhan:1997iv} for the
treatment of the phase shift. 

We start from the form of subtraction
\begin{align}
    \frac{1}{L}\sum_{n=-\infty}^\infty f(q_n) - \frac{1}{L}\sum_{n=-\infty}^\infty f(k_n) ,
    \label{eq:MNC1}
\end{align}
for an even function $f(x)$, where $q_n$ and $k_n$ are the discretized momenta in the kink and vacuum sectors in Eqs.~(\ref{eq:q_n}) and (\ref{eq:k_n}).
Using $f(x)=f(-x)$, Eq.~(\ref{eq:MNC1}) is rewritten as
\begin{align}
    \frac2L \sum_{n=0}^\infty f(q_n) - \frac2L \sum_{n=0}^\infty f(k_n) .
    \label{eq:MNC2}
\end{align}
Note that $q_n=-q_{-n-1}$ and $k_n=-k_{-n-1}$ from Eqs.~(\ref{eq:q_n}) and~(\ref{eq:k_n}).

To perform the subtraction in Eq.~(\ref{eq:MNC2}), one has to introduce the upper limits of the two sums to make them finite, and then take them to infinity keeping the difference finite. In the MNC, this cutoff is introduced in such a way that the mode numbers in the kink and vacuum sectors are equivalent. This prescription is justified, for example, in the lattice regularization. Then, since the kink sector has two bound states that are not included in Eq.~(\ref{eq:MNC2}), and there are two modes for each $n$ in Eq.~(\ref{eq:MNC2}), the upper bound for the kink sector is one smaller than the vacuum sector. Therefore, in the MNC the sum~(\ref{eq:MNC2}) is defined as
\begin{align}
    \frac2L \sum_{n=0}^{N-1} f(q_n) - \frac2L \sum_{n=0}^N f(k_n)
    = \frac2L \sum_{n=0}^{N-1} \Big( f(q_n) - f(k_{n+1}) \Big) - \frac2L f(k_0) ,
    \label{eq:MNC3}
\end{align}
where $N$ is half the number of the modes that are taken infinity at the end of the calculation. On the right-hand side of Eq.~(\ref{eq:MNC3}), the subtraction is taken between the modes off by one as in Fig.~\ref{fig:MNC}, and the remaining $n=0$ mode in the vacuum sector is put outside the sum.

From Eqs.~(\ref{eq:q_n}) and~(\ref{eq:k_n}), one sees that
\begin{align}
    q_n 
    = k_{n+1} - \frac{2\pi + \delta_{p}(q_n)}L
    = k_{n+1} - \frac{2\pi + \delta_{p}(k_{n+1})}L + {\cal O}(L^{-2}).
\end{align}
In the limit $L\to\infty$ we thus have 
\begin{align}
    f(q_n) - f(k_{n+1}) = -f'(k_{n+1}) \frac{2\pi + \delta_{p}(k_{n+1})}L.
    \label{eq:fq-fk}
\end{align}
Plugging Eq.~(\ref{eq:fq-fk}) into Eq.~(\ref{eq:MNC3}), Eq.~(\ref{eq:MNC1}) is calculated to be
\begin{align}
    &\frac{1}{L}\sum_{n=-\infty}^\infty  f(q_n) - \frac{1}{L}\sum_{n=-\infty}^\infty  f(k_n)
    \notag \\
    &=-\frac{2}{L}\lim_{N\to\infty} \sum_{n=0}^{N-1} f'(k_{n+1}) \frac{2\pi + \delta_{p}(k_{n+1})}L - \frac2L f(k_{0})
    \notag \\
    &\xrightarrow[L\to\infty]{} -\frac{2}{L}\int_0^\infty \frac{dk}{2\pi} f'(k)  ( \delta_{p}(k)+2\pi) - \frac2L f(k_{0})
    \notag \\
    &= -\frac{1}{\pi L}\Big[ f(k)( \delta_{p}(k)+2\pi)\Big]_0^\infty +\frac{2}{L} \int_0^\infty \frac{dk}{2\pi} f(k) \delta'(k) - \frac2L f(k_{0})
    \notag \\
    &= - \frac{3\sqrt{2}}{L\pi} \lim_{k\to\infty} \Big( f(k) \frac mk \Big)
    + \frac2L \int_0^\infty \frac{dk}{2\pi} f(k) \delta_p'(k) ,
    \label{eq:MNC4}
\end{align}
where in the last equality we used Eq.~(\ref{eq:delta_p->}).
The last term in Eq.~(\ref{eq:MNC4}) cancels out with $\delta_p'(q_n)/L$ term in Eqs.~(\ref{eq:V1})--(\ref{eq:V2}).
Substituting $f(x)=\sqrt{x^2+2m^2}$ and $f(x)=1/\sqrt{x^2+2m^2}$ into this result gives Eqs.~(\ref{eq:V1})--(\ref{eq:V2}), respectively.

In the above derivation, we imposed the APBC for both the kink and vacuum sectors. Other BCs have been employed in the literature for the calculation of the total energy of the kink~\cite{Dashen:1974cj,Rajaraman:1982is,Rebhan:1997iv}. These analyses have shown that the final result does not depend on the BCs employed in these studies. In particular, it has been pointed out that different BCs for the kink and vacuum sectors, for example, APBC for the kink sector and PBC for the vacuum sector, yield the same result~\cite{Goldhaber:2000ab}. Upon our examination, these arguments directly apply to our manipulation as well. Therefore, we expect that Eqs.~(\ref{eq:V1})--(\ref{eq:V2}) are valid even for other BCs.

\section{Calculations of $H_i(x)$}
\label{app:tadpole}

In this Appendix, we calculate $H_i(x)$ in Eq.~(\ref{eq:H_i}) that appear in the analysis of the tadpole diagram in Sec.~\ref{sec:tad}. 

We start from an identity~\cite{Gervais:1975pa}\footnote{See, Eq.~(4.7) of Ref.~\cite{Gervais:1975pa}.},
\begin{align}
    \int dy \partial^2_y \bar\phi_{\rm kink}(y)\tilde{G}(x,y)= \frac{i}{2}x \partial_x \bar\phi_{\rm kink}(x) .
    \label{eq:Gervais}
\end{align}
Substituting the EoM
\begin{align}
    (\partial^2_y +m^2 )\bar\phi_{\rm kink}(y) &= m^2 \bar\phi^3_{\rm kink}(y) ,
    \label{eq:eombar}
\end{align}
into Eq.~(\ref{eq:Gervais}), we obtain
\begin{align}
    \int dy \partial^2_y \bar\phi_{\rm kink}(y)\tilde{G}(x,y)&= m^2 \int dy (\bar\phi^3_{\rm kink}(y)-\bar\phi_{\rm kink}(y)) \tilde{G}(x,y) \notag \\
    &= m^2(H_3(x)-H_1(x)).
\label{eq:H3-H1}
\end{align}

By the partial integral, Eq.~(\ref{eq:Gervais}) is also calculated to be
\begin{align}
\int dy \partial^2_y \bar{\phi}_{\rm kink}(y)\tilde{G}(x,y)
=& \big[\partial_y \bar{\phi}_{\rm kink}(y)\tilde{G}(x,y) \big]_{-L/2}^{L/2}
-\int dy \partial_y \bar{\phi}_{\rm kink}(y)\partial_y \tilde{G}(x,y) 
\notag \\
=& -\big[ \bar{\phi}_{\rm kink}(y)\partial_y \tilde{G}(x,y) \big]_{-L/2}^{L/2}
+\int dy \bar{\phi}_{\rm kink}(y)\partial^2_y \tilde{G}(x,y) .
\label{eq:[]int}
\end{align}
Provided that the surface terms in the second and third lines vanish, one obtains
\begin{align}
\int dy \partial^2_y \bar{\phi}_{\rm kink}(y)\tilde{G}(x,y)
=\int dy \bar{\phi}_{\rm kink}(y)\partial^2_y \tilde{G}(x,y) .
\label{eq:[]int2}
\end{align}
Here, we note that the surface terms in Eq.~(\ref{eq:[]int}) vanish trivially when the APBC (or Dirichlet) BC is imposed on $\tilde\eta(x)$. We employ the APBC from this cancellation, although the surface terms would vanish even for other BCs because of $\lim_{x\to\pm\infty}\partial_x\bar\phi(x)=0$ and $\lim_{y\to\pm\infty}G(x,y)=0$.

Plugging
\begin{align}
    (-\partial^2_y -m^2+3 m^2 \bar\phi^2_{\rm kink}(y))\tilde{G}(x,y)&=-i\delta(x-y) ,
    \label{eq:G=delta}
\end{align}
into Eq.~(\ref{eq:[]int2}) gives
\begin{align}
\int dy \partial^2_y \bar{\phi}_{\rm kink}(y)\tilde{G}(x,y)
&= \int dy \bar{\phi}_{\rm kink}(y)\big \{ (-m^2+3m^2\bar{\phi}^2_{\rm kink}(y))\tilde{G}(x,y)+i\delta(x-y)  \big \} \notag \\
&=m^2(3H_3(x)-H_1(x))+i\bar{\phi}_{\rm kink}(x) .
\label{eq:3H3-H1}
\end{align}
From Eqs.~(\ref{eq:H3-H1}) and~(\ref{eq:3H3-H1}), one finds
\begin{align}
H_1(x)&=-\frac{i}{2m^2}\partial_x \big (x \bar{\phi}_{\rm kink}(x) \big ) ,\\
H_3(x)&=-\frac{i}{2m^2} \bar{\phi}_{\rm kink}(x)  .
\end{align}

To calculate $H_5(x)$, we use the following relations
\begin{align}
\int dy \bar{\phi}^3_{\rm kink}(y) \partial^2_y \tilde{G}(x,y)& =\int dy \bar{\phi}^3_{\rm kink}(y)\big \{ (-m^2+3m^2\bar{\phi}^2_{\rm kink}(y))\tilde{G}(x,y)+i\delta(x-y)  \big \} \notag \\
& =m^2(3H_5(x)-H_3(x))+i\bar{\phi}^3_{\rm kink}(x) , \\
\int dy \bar{\phi}^3_{\rm kink}(y) \partial^2_y \tilde{G}(x,y)&= \int dy \partial^2_y \big (\bar{\phi}^3_{\rm kink}(y) \big )  \tilde{G}(x,y) \notag \\
& =m^2(6H_5(x)-9H_3(x)+3H_1(x)) ,
\end{align}
which lead to 
\begin{align}
H_5(x)&=\frac{i}{2m^2} \partial_x \big (x \bar{\phi}_{\rm kink}(x) \big ) -\frac{4i}{3m^2} \bar{\phi}_{\rm kink}(x) +\frac{i}{3m^2} \bar{\phi}^3_{\rm kink}(x) .
\end{align}

\section{Other approaches}
\label{sec:LMR}

The energy densities obtained in Refs.~\cite{Goldhaber:2001rp,Martin:2022pri} and our result on $\braket{T^{00}(x)}$ differ with one another. In this Appendix, in order to gain insights into the origin of the difference we give a brief review of the regularization employed in Ref.~\cite{Goldhaber:2001rp} called the local-mode regularization (LMR).

In the LMR, vacuum subtraction is performed in an infinitely-long system. Since the eigenfunctions Eq.~(\ref{eq:psiq}) form a continuous spectrum in this case, it is convenient to use the orthogonality condition of eigenfunctions
\begin{align}
    \int_{-\infty}^{\infty}dx \check\psi_0(x) \check\psi_0(x) = 
    \int_{-\infty}^{\infty}dx \check\psi_1(x) \check\psi_1(x) = 1, \quad
    \int_{-\infty}^{\infty}dx \check\psi_{q}^*(x) \check\psi_{p}(x) = \delta(q-p),
\end{align}
in place of Eq.~(\ref{eq:ortho}).

According to Ref.~\cite{Goldhaber:2001rp}, the LMR introduces the local mode density
\begin{align}
\rho_{\Lambda}^{\rm K}(x)\equiv \sum^{N}_{l=0}\check{\psi}^*_{l}(x)\check{\psi}_{l}(x) 
=\check{\psi}^2_{0}(x)+\check{\psi}^2_{1}(x)+2\int^{\Lambda}_{0}\frac{dq}{2\pi}|\check{\psi}_{q}(x)|^2 ,
\label{eq:rhoK}
\end{align}
for the kink sector with a cutoff $\Lambda=2\pi N / L $ and the corresponding one $\rho_\Lambda^{\rm V}(x)$ for the vacuum sector
\begin{align}
    \rho^{\rm V}_\Lambda(x) = 2\int^\Lambda_{0}\frac{dk}{2\pi}.
    \label{eq:rhoV}
\end{align}

Then, the UV cutoff in each sector, $\Lambda_{\rm K}$ and $\Lambda_{\rm V}$, which are dependent on $x$, is introduced so that the local mode densities are equivalent for each sector
\begin{align}
\rho^{\rm K}_{\Lambda_{\rm K}}(x) &=\rho^{\rm V}_{\Lambda_{\rm V}}(x) ,
\label{eq:rhoK=V}
\\
\Lambda_{\rm V} &= \Lambda_{\rm K}+ \Delta\Lambda(x).
\end{align}
Using the completeness relation
\begin{align}
\int^{\infty}_{-\infty}\frac{dq}{2\pi}\{|\check{\psi}_{q}|^2-1 \}+\check{\psi}^2_{0}+\check{\psi}^2_{1}=0,
\end{align}
Eq.~(\ref{eq:rhoK}) is given by 
\begin{align}
\rho^{\rm K}_{\Lambda} = - 2\int^{\infty}_{\Lambda} \frac{dq}{2\pi}| \check{\psi}_{q}|^2 +  2 \int^{\infty}_{0} \frac{dq}{2\pi} .
\label{eq:rhoKint}
\end{align}
From Eqs.~(\ref{eq:rhoKint}) and~(\ref{eq:rhoV}) one obtains
\begin{align}
\Delta \Lambda (x)
&=\int^{\infty}_{\Lambda}dq\big\{ 1-|\check{\psi}_{q}|^2  \big \} 
\notag \\
&=\int^{\infty}_{\Lambda}dk \Big\{ \frac{3}{2(k^2+2m^2)} \psi^2_0 +\frac{3}{2k^2+m^2} \psi^2_1\Big \}
= \frac{3m^2}{2\Lambda} (\psi^2_0+\psi^2_1) + \mathcal{O}(\Lambda^{-2})\notag \\
&= \frac{3m^2}{2\Lambda} {\rm sech}^2 \frac{mx}{\sqrt{2}} + \mathcal{O}(\Lambda^{-2}) .
\label{eq:DeltaL}
\end{align}

Using Eq.~(\ref{eq:DeltaL}), the vacuum subtractions in Eqs.~(\ref{eq:V1}) and~(\ref{eq:V2}) are calculated to be
\begin{align}
&\frac{1}{2}\int^{\infty}_{-\infty}\frac{dq}{2\pi} \sqrt{q^2+2m^2}-\frac{1}{2}\int^{\infty}_{-\infty}\frac{dk}{2\pi}  \sqrt{k^2+2m^2} \notag \\
&=\int^{\Lambda}_{0}\frac{dq}{2\pi} \sqrt{q^2+2m^2}-\int^{\Lambda+\Delta \Lambda}_{0}\frac{dk}{2\pi} \sqrt{k^2+2m^2} \notag \\
&=-\int^{\Lambda+\Delta \Lambda}_{\Lambda}\frac{dk}{2\pi} \sqrt{k^2+2m^2}
=-\frac{\Lambda\Delta \Lambda }{2\pi}+ \mathcal{O}(\Lambda^{-1}) \notag \\
&\xrightarrow[\Lambda \to \infty]{}-\frac{3m^2}{4\pi} {\rm sech}^2 \frac{mx}{\sqrt{2}} ,
\label{eq:LMR1}
\\
&\frac{1}{2}\int^{\infty}_{-\infty}\frac{dq}{2\pi} \frac{1}{\sqrt{q^2+2m^2}}-\frac{1}{2}\int^{\infty}_{-\infty}\frac{dk}{2\pi}  \frac{1}{\sqrt{k^2+2m^2}} \notag \\
&=\frac{\Delta \Lambda }{2\pi\Lambda}+ \mathcal{O}(\Lambda^{-3}) 
\xrightarrow[\Lambda \to \infty]{}0  ,
\label{eq:LMR2}
\end{align}
Since the LMR is needed only for the subtraction between divergent sums, the replacement of Eqs.~(\ref{eq:V1}) and~(\ref{eq:V2}) with Eqs.~(\ref{eq:LMR1}) and~(\ref{eq:LMR2}), respectively, is only the change in the LMR compared with the MNC. 
Thus, the final result of the EMT distribution in the LMR is obtained by simply replacing
\begin{align}
-\frac{3\sqrt2m}{2\pi L} \longrightarrow -\frac{3m^2}{4\pi}{\rm sech}^2 \frac{mx}{\sqrt{2}} ,
\label{eq:->LMR}
\end{align}
in Eqs.~(\ref{eq:T00final}) and~(\ref{eq:T11final}).
This result gives $\Delta T_{\rm kink}^{11}(x)=-3m^2/(4\pi) {\rm sech}^2(mx/\sqrt{2})$, which is not consistent with the momentum conservation $\partial_1 T^{11}=0$, while the spatial integral of $\braket{\tilde{T}^{00}(x)}$ reproduces the result in Ref.~\cite{Dashen:1974cj} even after the replacement.

We, however, note that the result of $\Delta T_{\rm kink}^{00}(x)$ obtained with the replacement~(\ref{eq:->LMR}) does not agree with the energy density in Ref.~\cite{Goldhaber:2001rp}. This suggests the existence of a difference in the manipulation other than the vacuum subtraction scheme. On the other hand, we found that this result agrees with the energy density in Ref.~\cite{Martin:2022pri}, while the point-split regularization is employed there. The agreement implies the similarity of the regularization schemes. We, however, do not pursue details further in the present study.

\bibliographystyle{JHEP}
\bibliography{phi4kinkEMT}
\end{document}